\pdfoutput=1
\documentclass[%
 reprint,
superscriptaddress,
 amsmath,amssymb,
 aps,
]{revtex4-2}
\usepackage[english]{babel}
\makeatletter
\AtBeginDocument{%
  \renewcommand{\selectlanguage}[1]{%
    \def\@tempa{#1}%
    \def\@tempb{en}\def\@tempc{EN}\def\@tempd{english}%
    \ifx\@tempa\@tempb\relax
    \else\ifx\@tempa\@tempc\relax
    \else\ifx\@tempa\@tempd\relax
    \else\PackageWarning{babel}{Unknown language `#1' in bibliography; ignored}%
    \fi\fi\fi
  }%
}
\makeatother
\usepackage[dvipsnames,table,xcdraw]{xcolor}
\usepackage{graphicx}
\usepackage{textcomp}
\usepackage{capt-of}
\usepackage{placeins}
\usepackage{hyperref}
\hypersetup{
    colorlinks=true,
    linkcolor=blue,
    filecolor=magenta,
    urlcolor=blue,
}

\usepackage{amsfonts}
\usepackage{amsmath}
\usepackage{amssymb}
\usepackage{braket}
\usepackage{tikz}
\usetikzlibrary{arrows.meta, positioning, shapes.geometric}
\usepackage{enumerate}
\usepackage[normalem]{ulem}
\usepackage{soul}
\usepackage{multirow}
\usepackage{rotating}
\usepackage{verbatim}
\usepackage{bm}
\DeclareMathOperator{\Tr}{Tr}

\usepackage[sort&compress]{natbib}

\newcommand{\startsupplnote}{\clearpage}
\newcounter{supplnote}
\newcommand{\supplnote}[2]{%
  \startsupplnote
  \refstepcounter{supplnote}%
  \edef\supplnotenum{\thesupplnote}%
  \subsection*{Supplementary Note~\supplnotenum: #2}%
  \label{#1}%
}
\newbox\sibibbox

\begin{document}
%\linenumbers

\title{Broadband Control of Light through Complex Media via Automatic Self-Referencing Transmission Matrix Characterisation}

%First
\author{Marcos Maestre Morote}
\thanks{Corresponding author; E-mail: m.maestremorote@uq.edu.au}
\affiliation{School of Electrical Engineering and Computer Science, The University of Queensland, Brisbane, QLD 4072, Australia}
%Others
\author{Mickael Mounaix}
\affiliation{School of Electrical Engineering and Computer Science, The University of Queensland, Brisbane, QLD 4072, Australia}

\author{Andrew Komonen}
\affiliation{School of Electrical Engineering and Computer Science, The University of Queensland, Brisbane, QLD 4072, Australia}

%Last authors
\author{Martin Pl\"{o}schner}
\thanks{These authors share last authorship.}
\affiliation{School of Electrical Engineering and Computer Science, The University of Queensland, Brisbane, QLD 4072, Australia}

\author{Joel Carpenter}
\thanks{These authors share last authorship.}
\affiliation{School of Electrical Engineering and Computer Science, The University of Queensland, Brisbane, QLD 4072, Australia}

\begin{abstract}

\textbf{Light propagation through complex media underpins critical optical technologies, from imaging distant stars and ground-to-space communication to imaging inside biological tissue with hair-thin multimode fibre endoscopes. Central to controlling light through such disordered media lies the transmission matrix. Inaccessible to accurate modelling, the transmission matrix must be measured experimentally---yet this conventionally relies on techniques involving an external phase reference. For low-coherence or broadband sources, the stringent coherence, mode-matching, temporal-overlap, and stability requirements of that external reference can make such characterisation prohibitively difficult or fundamentally infeasible. Alternatively, existing self-referencing techniques use algorithmically fragile global optimisation methods, relying on fixed preselected internal reference(s), whose incomplete overlap with the transmitted field can create measurement blind spots. Here, we introduce an automatic self-referencing measurement technique based on spatial state tomography that circumvents these coherence and algorithmic limitations. Rather than relying on a preselected reference or complex phase retrieval, our approach systematically leverages the local interference among all propagating modes as distributed phase references without prior assumptions. We demonstrate this framework experimentally for a multimode fibre across coherent, low-coherence, and broadband regimes, recovering its complete optical transmission matrix and performing high-fidelity spatial and polarisation beam shaping in each case. By enabling robust, source-matched, self-referencing transmission-matrix measurement, our method extends light control through complex media into broadband illumination regimes relevant to biomedical imaging, optical communications, and high-power laser applications.}

\end{abstract}

\maketitle

% Flag so paper.tex skips its standalone document wrapper when included here.
\newcommand{\paperloadedbymain}{}
% !TEX root = main.tex
% Standalone wrapper for IDE/texlab when this file is checked alone.
% When \input from main.tex, \paperloadedbymain is defined and this is skipped.
\ifdefined\paperloadedbymain
\else
\documentclass[reprint,superscriptaddress,amsmath,amssymb,aps]{revtex4-2}
\usepackage[english]{babel}
\usepackage{graphicx,braket,amsmath,amssymb,bm,natbib,hyperref,xcolor,tikz,enumerate,ulem,soul,multirow,rotating,verbatim,textcomp,capt-of,amsfonts}
\begin{document}
\fi
\section*{INTRODUCTION}

Characterising how light undergoes multiple scattering in complex media — from plasmonic systems and nanophotonic structures~\cite{anttuScatteringMatrixMethod2011,gopinathPhotonicPlasmonicScatteringResonances2008} to highly turbid biological tissues~\cite{yaqoob_optical_2008,jacques_optical_2013} and multimode fibres (MMF)~\cite{ploschner_seeing_2015,cizmar_exploiting_2012}---has long been a formidable challenge. At the heart of this difficulty is the seemingly random scrambling of optical waves, which obscures spatial information and limits the ability to transmit or recover structured optical fields, preventing, for example, the formation of clear images through these systems. Despite this apparent complexity, the underlying scattering process remains linear and deterministic and can therefore be fully described by a transmission matrix (TM), which maps the optical field from a defined input plane to an output plane~\cite{popoff_measuring_2010,vellekoop_focusing_2007}. By capturing this input–output relationship, the TM provides a powerful framework for characterising and controlling scattering systems that would otherwise appear intractable.

The optical transmission matrix (TM) is a complex, wavelength-dependent matrix that describes how each input spatial and polarisation mode couples to every supported output mode at each wavelength ($\lambda$). Predicting the TM theoretically is generally impractical---not only because modelling multiple scattering can be computationally demanding, but more importantly because the microscopic structure and experimental state of the scattering system are rarely known with sufficient accuracy. The TM must therefore typically be measured experimentally by sequentially launching a set of input fields at the proximal end using a spatial light modulator (SLM), such as an LCoS or DMD, or a raster-scanning device such as a steerable mirror. The corresponding complex output fields at the distal end are then recovered using techniques such as off-axis digital holography~\cite{kreis_methods_1997, mounaix_temporal_2017, carpenter_digholo_2022, aulbach_control_2011, mounaix_spatiotemporal_2016} or MIMO processing~\cite{ryf_space-division_2011, fontaine_space-division_2012}. Among these approaches, digital holography (DH) is widely regarded as the gold standard because it measures the complex amplitude of each output field relative to an external, ideally well-defined reference derived from the same source as the input fields. For a long-coherence source (LCS), or a sufficiently phase-stable pulsed source, such a reference can be realised, although at the cost of constructing and stabilising an interferometer. For short-coherence or broadband sources, however, the requirements become considerably more stringent: the reference must remain mutually coherent with the transmitted field, be spatially and polarisation matched, overlap temporally within the short coherence length, and maintain wavelength-scale path stability. Any mismatch reduces interference visibility or leaves parts of the transmitted field insufficiently referenced, compromising recovery of the complete TM. Consequently, a suitable reference is not merely inconvenient to provide but may be impractical or fundamentally impossible to realise for sources with short coherence lengths, broad or structured spectra, unstable emission, or poorly defined spatial and polarisation states. The problem is further compounded under low-light conditions, including photon-starved and single-photon measurements, where dividing the available signal between reference and measurement arms, together with weak interference visibility and detector noise, can make reference-based recovery prohibitively noisy and prevent accurate reconstruction of the complete TM.

These limitations have motivated self-referenced approaches that recover the output field without routing an external reference. This implies using an internal transmitted field, either directly or indirectly, to recover its own phase. Correlation filters methods provide relatively simple implementation~\cite{carpenter_110x110_2014, carpenter_degenerate_2012}. However, they incur beamsplitter-like losses that scale with the number of spatial modes, decreasing the signal-to-noise (SNR), and require a complicated device such as an SLM at the distal detection end. Such losses, become especially problematic in highly multimode systems and low photon count measurements. Therefore, self-referencing power-efficient and cost-effective approaches that are applicable to both narrowband and broadband light sources, as well as simplifying both construction and calibration procedures at the detection side are of vital importance.

Addressing this need, most single-camera, intensity-only methods that recover the complete TM without an external reference follow one of two strategies. The first uses a specific output field, such as a superposition of modes, as a static phase reference~\cite{popoff_image_2010}. This can be suitable for some open scattering systems with a nearly infinite number of supported modes, where the pre-selected reference field is likely to be distributed across the detection plane. However, this strategy fails at locations where the reference field is weak or absent. These reference-dependent blind spots are particularly severe in MMFs, which support a finite number of guided modes and can therefore produce highly structured output fields with limited overlap between individual modal components. Overcoming this challenge is critical, as MMFs are pivotal enablers in several industries, including telecommunications~\cite{richardson_space-division_2013, rademacher_153_2022}, biomedical imaging~\cite{mahalati_resolution_2013, ploschner_multimode_2015}, and high-power laser systems~\cite{jauregui_transverse_2020, chen_mitigating_2023}. Furthermore, the low absorption and comparatively well-defined modal structure of MMFs make them an excellent testbed for studying nonlinear and quantum phenomena, including multimode solitons and photon entanglement~\cite{wright_physics_2022, lib_quantum_2022}. The second strategy avoids a fixed reference by applying many random input projections and recording only the corresponding output intensities. Individual rows, or sometimes the full TM, are then inferred using optimisation or phase-retrieval algorithms~\cite{dremeau_reference-less_2015, ngom_mode_2018, wang_feedback-assisted_2021}. Although this removes the need to select a particular co-propagating reference, it replaces an experimental referencing problem with a computational inverse problem. The reconstruction is generally non-convex, computationally demanding, sensitive to noise and initial conditions, and susceptible to stagnation in local minima, so convergence to the complete physical TM cannot be guaranteed. Machine-learning-based single-camera approaches have also been proposed~\cite{li_speckle_2022,wang_deep_2018,yolalmaz_hyper-spectral_2023, goel_referenceless_2023}, but these similarly require large training datasets and optimisation of a model whose accuracy depends on the representativeness and stability of the training conditions. Moreover, both conventional phase-retrieval and machine-learning approaches generally rely on stable, high-contrast speckle patterns produced by mutually coherent modal interference. Their applicability therefore deteriorates for broadband or short-coherence sources, for which wavelength-dependent and temporally incoherent contributions wash out the interference information on which the reconstruction depends.

Existing methods therefore offer only partial solutions. Static self-references retain reference-dependent blind spots, correlation-filter methods sacrifice optical throughput as the modal dimensionality increases, and intensity-only approaches rely on computationally heavy and potentially ambiguous optimisation. A complete solution requires a measurement that avoids an external reference without selecting a privileged internal reference channel and preserves the available photon budget. It must directly capture the required complex modal relationships without iterative phase retrieval and remain applicable to both long- and short-coherence sources.
 
In this work, we harness spatial-state tomography (SST)~\cite{toninelli_concepts_2019, agnew_tomography_2011, ploschner_spatial_2022, dahl_high-dimensional_2021,carpenter_optical_2020} to retrieve the TM of an MMF using only intensity measurements at the distal fibre output, with the recovered matrix inherently matched to the coherence properties of the employed light source. Within the SST formalism, we launch a set of analyser states $\ket{\Omega_k}$ ($k \in 2N^2-N$) at the proximal fibre input and use the resulting intensity measurements to construct a high-dimensional Stokes vector $\mathbf{S}_{x,y}$ at each output-camera pixel $(x,y)$. Each pixel-dependent Stokes vector reveals the complex superposition of $N$ input modes, expressed in the basis of choice ($\{ \ket{\psi_n} \}$), required to focus light onto that pixel, thereby recovering the optical TM up to an unknown phase factor $\Delta\phi_{x,y}$ associated with each output position. These relative phase factors may appear inaccessible because different camera pixels do not interfere directly in the detection plane. However, we show that they can be reliably recovered when multiple camera pixels sample each modal speckle grain, leveraging the interference between spatial modes at the proximal side of the fibre. The resulting phase-corrected TM is then transformed into an $N\times N$ mode transmission matrix (MTM) expressed in the basis of the fibre's ideal theoretical eigenmodes. Finally, we validate the method by retrieving the fibre MTM and using it for distal beam shaping with both narrowband long-coherence sources (LCSs) and broadband short-coherence sources (SCSs).

\section*{RESULTS}

\begin{figure*}[ht!]
    \centering
    \includegraphics[width=0.99\linewidth]{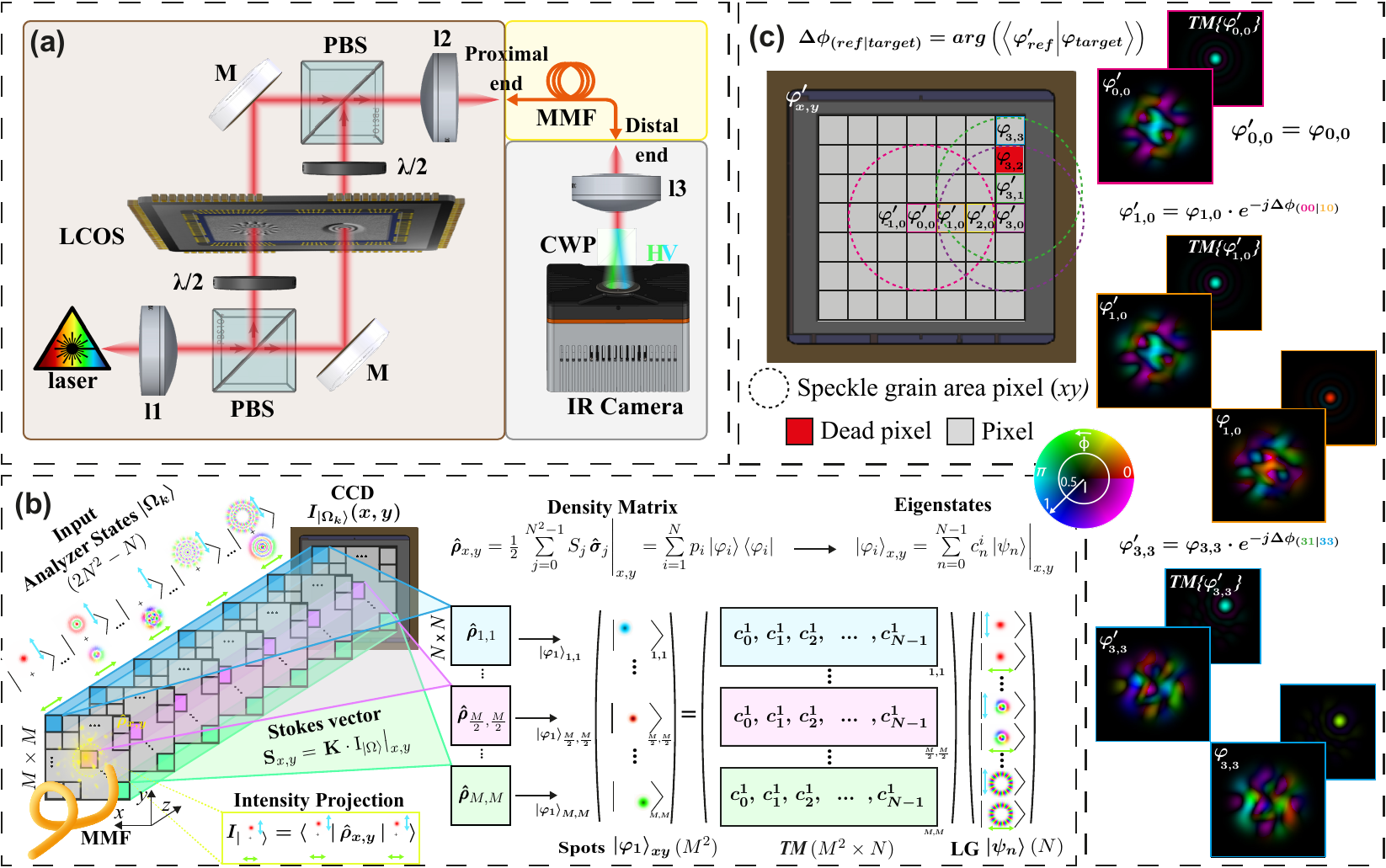}
    \caption{Optical setup, SST and phase locking working principle. (a) Optical setup composed of three color-coded stages. (a.brown) Wavefront shaping stage comprising lenses, polarisation beam splitters (PBS), and half-wave plates to provide independent horizontal (H) and vertical (V) polarisation control. (a.yellow) MMF under test. (a.gray) Detection stage formed by a collimation lens, calcite Wollaston polariser (CWP) and a camera for parallelisation of each pixel intensity measurement. (b) SST pipeline processing scheme. The intensity value per pixel ($I_{x,y}$) for each input analyser state $\ket{\Omega_k}$ is processed into a density matrix ($\hat{\rho}_{x,y}$). Finally, the relationship between input and output basis is established by extracting the first eigenstate of each $\hat{\rho}_{x,y} \rightarrow \ket{\varphi_1}$. (c) Iterative phase locking algorithm where the phase of each spot/pixel, expressed as complex vector of coefficients ($\varphi_{x,y} = \{c_1, c_2 ... c_n\}$) is measured with respect to a global phase, in this case, the centre pixel. The unknown pixel's phase-offset is extracted by overlapping phase-corrected neighbouring pixels ($\varphi'_{x,y}$) inside the same speckle grain.}
    \label{fig:figure1}  
\end{figure*}

The fundamental principle of SST is to frame complex-media characterisation as a high-dimensional spatial analogue of Stokes polarimetry per output-camera pixel, where each input-to-output channel relationship is encoded as an $N\times N$ density matrix $\hat{\rho}_{x,y}$ in the fibre's $N$-mode Hilbert space. Just as conventional polarimetry reconstructs a polarisation state from intensity projections onto the Pauli operator basis---thereby sampling all directions on the Poincaré sphere---SST reconstructs the $\hat{\rho}_{x,y}$ from intensity projections onto the high-dimensional generalisation of that basis formed by the Gell--Mann matrices ($\hat{\sigma}$), (Supplementary Note~\ref{sn:sst-theory}). This tomographically complete measurement set eliminates the need for an external reference entirely, unlike conventional self-referenced techniques, in which a preselected co-propagating field can vanish at certain output locations and create reference-dependent ``dead zones''. SST, by design, forces every mode to interfere with every other mode, guaranteeing that if a suitable reference exists, it will be found.

This unguided, automatic self-referencing (Fig.~\ref{fig:figure1}) is realised by sequentially exciting a complete set of $2N^2-N$ analyser states $\ket{\Omega_k}$, at the proximal end of the fibre. These input fields are pairwise interference patterns between each mode combination at four equally spaced phase offsets, plus each mode by itself. Mathematically, the analyser states correspond to the unique eigenvectors of $\hat{\sigma}_j$, and their eigenvalues ($\kappa_j^r$) are used to construct a matrix operator $\mathbf{K}$, which acts as a weighting factor upon the measured intensity values, yielding the complete high-dimensional Stokes vector $\mathbf{S}_{x,y}$ at each pixel independently. The mode basis $\ket{\psi_n}$ used to construct these states may be any convenient orthogonal set spanning the full fibre mode space; for graded-index fibres with near-parabolic profiles, Laguerre--Gaussian (LG) or Hermite--Gaussian (HG) modes are a natural choice. In summary, SST addresses complex-media characterisation in two complementary steps. First, for each output pixel $(x,y)$, the eigenstates $\ket{\varphi}$ of $\hat{\rho}_{x,y}$ identify the input modal superposition that maximises power delivery to that channel, equivalently the focusing solution for that input-to-output link. Second, by recovering the relative phase between neighbouring pixels within the same speckle grain, these channel-resolved solutions are stitched into a globally, phase-corrected mode transmission matrix (MTM). This combination distinguishes SST from self-referenced approaches that may infer isolated input--output relationships but lack the cross-pixel phase consistency required to reconstruct the full complex matrix needed for arbitrary distal beam shaping.

\subsection*{Experimental setup}

We perform the SST measurement using the optical system illustrated in Fig.~\ref{fig:figure1}(a). It consists of a light source, an SLM system used to create the input analyser states, the fibre to be characterised, and a camera equipped with polarisation diversity optics to record the interference effect at each camera pixel. The light source is a swept-frequency laser (New Focus Venturi TLB-8800, $\lambda_c$ = 1300~nm, 80~nm tunning range) which serves as both, an LCS (narrowband-CW) and an emulated SCS (broadband) of tunable coherence lengths. The SLM based on an LCoS (Meadowlark P1920) is able to convert the input Gaussian beam from the light source into an arbitrary spatial/polarisation field using a Gerchberg-Saxton (GS) algorithm. The fibre-under-test is a 2~m length OM1 graded-index (GI) fibre (62.5~\textmu m core). The fibre supports 20 near-degenerate mode groups at 1300~nm yielding 420 spatial/polarisation modes. The distal facet of the fibre is imaged onto a Xenics Cheetah camera with a resolution of 640$\times$512 pixels and 20~\textmu m pixel pitch. The polarisation diversity is achieved via a Wollaston prism such that each polarisation (H, V) component can be observed on separate sides of the camera frame. Selecting LG basis as the natural basis of the target GI-MMF with $N = 420$ modes, a total of 352,380 input fields are generated by the SLM system, and the corresponding camera frames are captured. From each of the $k$-frame, we extract two regions of interest of 200$\times$200 pixels associated with a polarisation component (H and V), which is processed independently. For each $(x,y)$-pixel, there is a set of $k-$intensity measurements $I_{\ket{\Omega_k}}$, which can be processed to become a high-dimensional Stokes vector ($\mathbf{S}_{x,y}$) of dimension $N^2$ or its alternate density matrix representation ($\hat{\rho}_{x,y}$) of dimension $N\times N$, as illustrated in Fig.~\ref{fig:figure1}(b). In the case of spatially coherent measurements without any experimental inaccuracies, the density matrix has a single non-zero eigenstate ($\ket{\varphi_1}$), revealing the spatial input mode superposition that will maximise the power at a given output pixel on the camera. For spatially incoherent light, multiple non-zero eigenstates $\ket{\varphi_i}$ ($i \in N$) will arise since there is no longer a single spatially coherent solution which can focus all the light to a given pixel. The $i$-eigenvalue ($p_i$) quantifies the power delivered into the $i$-solution, while the $i$-eigenvector provides the needed modal input superposition.

At this stage, if the fibre is intended to be used for simple point scanning at the distal end applications, such as fibre imaging, no additional processing is necessary. However, if the full complex transfer matrix is to be recovered, the phase relationship between the points on the camera must be established. Initially, the task of recovering the phase relationship between points may appear impossible, given that these points have never physically interfered in the camera plane. As a result, there is no common phase reference to lock them. Nevertheless, if the pixels on the camera are at least four times smaller than the diffraction-limited spot size (speckle grain) of the supported fibre modes ($M^2 \geq 4N$), the input states necessary to focus on neighbouring pixels on the camera must be in phase. Hence, by selecting one of the obtained input eigenstates ($\varphi_{x,y}$) associated with a focusing spot at location $(x,y)$ as a starting reference (for instance the centre one), we can calculate the phase offset ($\Delta\phi_{ref|target}$) with adjacent pixels and impose that the phase should not have changed, Fig.~\ref{fig:figure1}(c). This approach is repeated sequentially with new corrected pixels ($\varphi'_{x,y}$) acting as new references. This logic is explained in more detail in Methods~\ref{meth:phase}. Despite the simplicity, the algorithm is robust against anomalous pixels that could lead to cascading phase errors; dead pixels can be omitted since any already phase-locked eigenstate is a valid reference when the next target pixel belongs to the same speckle grain. Pixel phase can therefore be recovered reliably even under noisy acquisition, as the phase linking depends on the eigenvectors, which encode the input focusing solutions, and these remain stable, whereas noise and insufficient temporal resolution primarily affect the eigenvalues, inflating the apparent rank of $\hat{\rho}_{x,y}$ through additional weak, spurious eigenstates. Selecting the dominant real eigenstates therefore retains the physical solution needed for phase recovery. Finally, the H-V matrices mapping $N$ LG-basis at the input to $M^2$ spot-basis at the output are combined into a single MTM of dimension $N \times N$ (Supplementary Note~\ref{sn:mmf-mtm}). As detailed in~\cite{ploschner_seeing_2015, mahalati_resolution_2013}, the condition $M^2 = 4N$ is sufficient to reliably retrieve the complete MTM. If there is a need to reduce processing time and/or data acquisition storage, downsampling can be applied.

\subsection*{Long coherence source experimental results}

\begin{figure*}[ht!]
    \centering
    \includegraphics[width=0.9\linewidth]{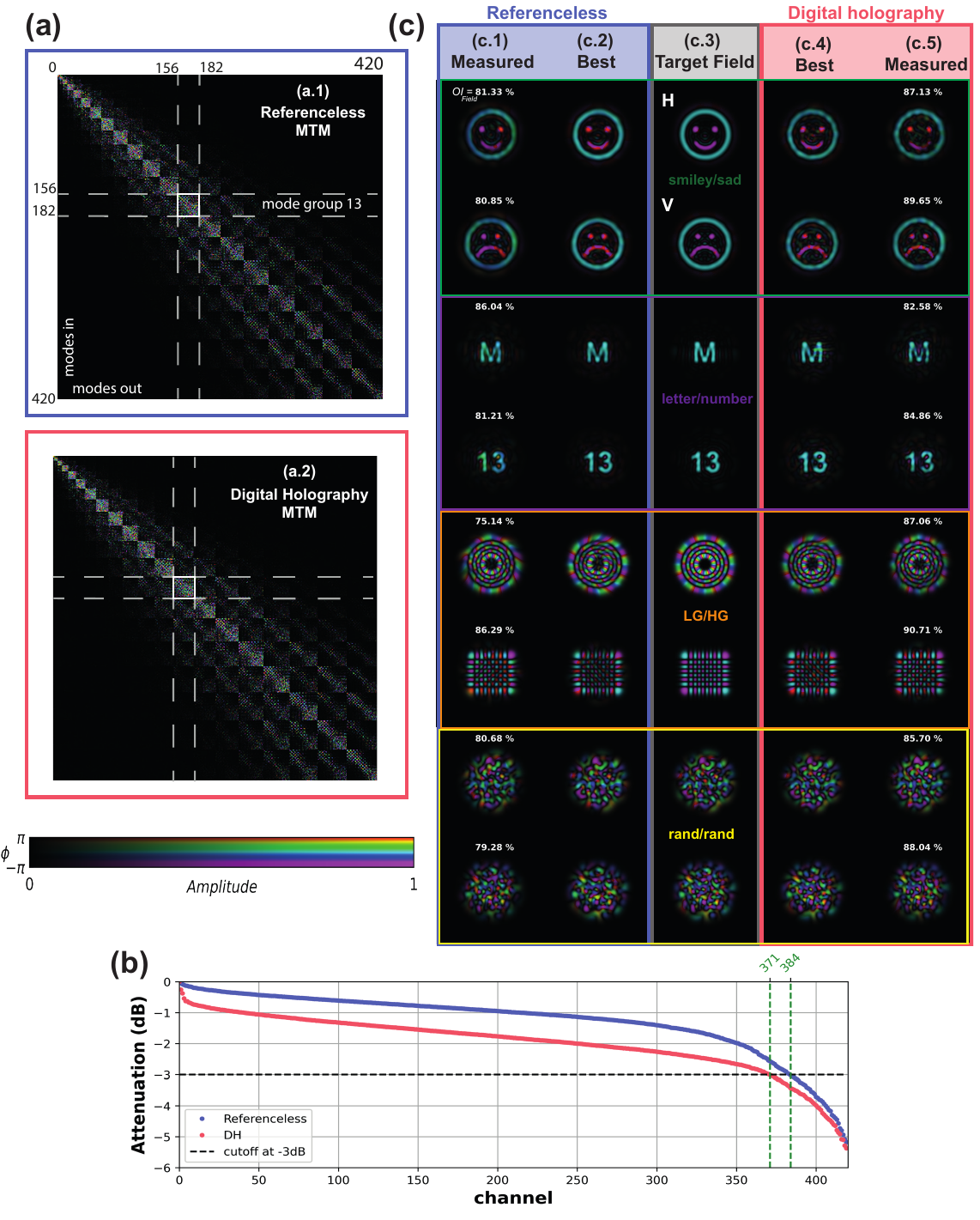}
    \caption{(a) Experimental (a.1) MTM$_{SST}$ retrieved using SST and (a.2) MTM$_{DH}$ measured with off-axis digital holography. (b) Normalised singular values of (a) matrices. (c) Target field (c.3), theoretical best (c.2,4), and experimentally measured fields (c.1,5) at the camera side measured using off-axis digital holography for MTM$_{SST}$ and MTM$_{DH}$ respectively. The fields overlap (in percentage) between the target fields and the experimental field is displayed on top of each pattern}
    \label{fig:figure2}
\end{figure*}

This scenario represents an entirely spatially coherent solution where the fibre is characterised at the laser centre wavelength  $\lambda_c = 1300$~nm. The SST measurement to retrieve the complete 420$\times$420 MTM$_{SST}$ associated with $\ket{\varphi_1}$ takes approximately 54~h due to the slow LCoS refresh rate (5~Hz) and on-the-fly hologram generation. The acquisition time could be drastically reduced with a faster SLM, ideally a DMD, and buffered pre-calculated holograms. The post-processing is CPU-GPU accelerated using an Intel CPU i9-9980XE@ 3~GHz equipped with an Nvidia Titan V GPU and takes around 30~min to extract an MTM$_{SST}$ from the 110~GB of raw data. The retrieved MTM$_{SST}$ is benchmarked against an MTM$_{DH}$ measured using off-axis digital holography right after the tomographic measurement (complete setup is illustrated in Supplementary Note~\ref{sn:lab-setup}). The DH method follows~\cite{mounaix_control_2019, carpenter_digholo_2022} and provides an independent ground truth against which to benchmark compare SST. The Fig.~\ref{fig:figure2}(a) illustrates both measured MTMs, whilst Fig.~\ref{fig:figure2}(b) is a comparison of the normalised singular values of each respective matrix. At first inspection, the MTM$_{SST}$ seems to provide a more accurate fibre description than MTM$_{DH}$ having more channels close to 0~dB attenuation before cutoff and hence higher spatial resolution. However, to definitively test this observation, we generate experimental complex fields at the fibre tip using the experimental MTM and calculating the overlap integral ($OI$) between the measured field ($\Vec{E}_{M}$) and the target field ($\Vec{E}_{T}$) as described in the Methods~\ref{meth:OI}. In order to show control of the 420 modes simultaneously (210 per polarisation), we have selected four different target fields with different difficulties, Fig.~\ref{fig:figure2}(c), where the most challenging one corresponds to the generation of a superposition of modes with arbitrary amplitudes and phases at both polarisations simultaneously.

In terms of field overlap, depicted on top of the measured fields, the MTM$_{DH}$ seems to perform better for the majority of the proposed patterns. Nevertheless, visual evaluation comparing measured and target fields seems to imply the opposite. An objective way to evaluate this subjective discrepancy in apparent quality is by an intensity overlap neglecting the phase as shown in Supplementary Note~\ref{sn:lcs-extra}, which validates the visual perception. This field overlap reduction could be caused by extra aberrations in the system, coming from the reference wave itself used to measure the fields and/or other optics that have not been digitally compensated for during the off-axis digital holography calibration. The other possible source of error could be induced by the long acquisition measurement time, where not only the fibre MTM drifts with temperature but also the mechanical components holding the fibre. The phase artefact is clear on top Fig.~\ref{fig:figure2}(c.3) smiley/sad and M/13, where a flat phase is expected. However, it disappears when a shorter measurement is taken where the temperature drift is negligible, as explored in Supplementary Note~\ref{sn:lcs-extra}. Finally, analysing the best possible fields that can be generated with the measured MTMs as shown in Fig.~\ref{fig:figure2}(c.2/4), MTM$_{SST}$ provides sharper patterns closer to the target, which correlates with the singular value analysis, as illustrated in Methods~\ref{meth:SVD}. This clearly indicates that the SST provides an accurate description of the fibre and the $OI$ reduction is a consequence of mechanical drifts with respect to the external reference to recover the phase. Although in some instances, our method outperformed off-axis digital holography and vice-versa, it is possible that a fairer comparison for the quality of each approach would involve averaging many digital holography measurements over a similar time period. Nonetheless, our self-referenced approach has two main advantages with respect to the holography technique: (1) the signal and the reference share a quasi-common path, (2) it effectively enforces that the average phase should be flat in the camera plane, meaning no optical aberrations present in the system between the distal fibre facet and the camera.

\subsection*{Short coherence source experimental results}

\begin{figure*}[ht!]
    \centering
    \includegraphics[width=0.99\linewidth]{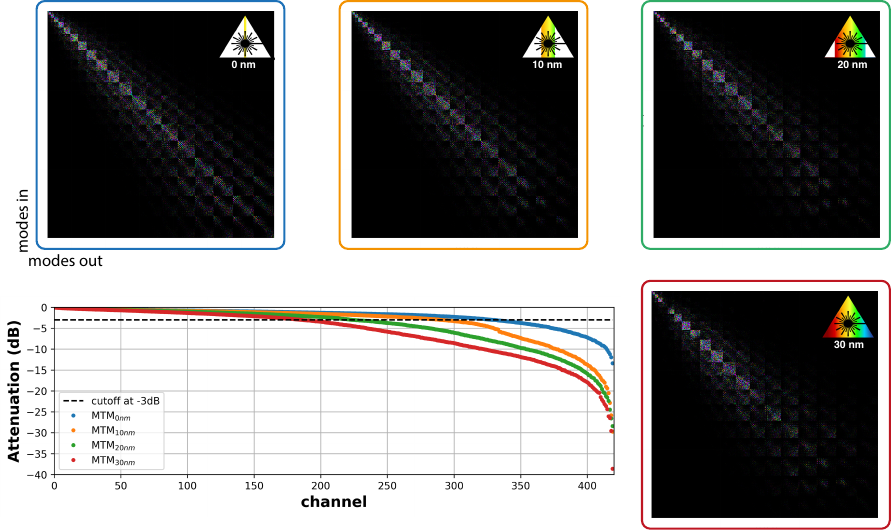}
    \caption{Mode transmission matrix analysis for different low coherent sources. Measured MTM$_{\Delta\lambda_S}$ color-coded as a function of the source bandwidth $\Delta\lambda_S$ for 0, 10, 20 and 30~nm, equivalently to 0, 1, 2 and 3 times the correlation bandwidth of the fibre ($\Lambda$). The singular values of each MTM$_{\Delta\lambda_S}$ provide the attenuation per channel. When the source $\Delta\lambda_S$ is in excess of several multiples of $\Lambda$, the number of simultaneous available eigenchannels is drastically reduced.}
    \label{fig:short_MTM_SVD}
\end{figure*}

To illustrate the applicability of our method to work with essentially any coherence length source, in the following scenario, the MMF is characterised for different source bandwidths ($\Delta\lambda_S$). Due to wavelength-dependent spatial mode dispersion, there is no single MTM$_{\lambda}$ which completely describes the propagation through the fibre over a sufficiently broad bandwidth range ($\Lambda$). Over this spectral bandwidth ($\Lambda < \Delta\lambda_S$), also referred to as the fibre correlation bandwidth, spatial decoherence will occur, resulting in multiple non-zero eigenstates. Nevertheless, SST will reveal which modes are mutually spatially coherent, assembling mutually incoherent MTM($\varphi$) which can deliver the maximum possible power to a desired $\varphi_i$ for a given source $\Delta\lambda_S$.

\begin{figure*}[ht!]
    \centering
    \includegraphics[width=0.99\linewidth]{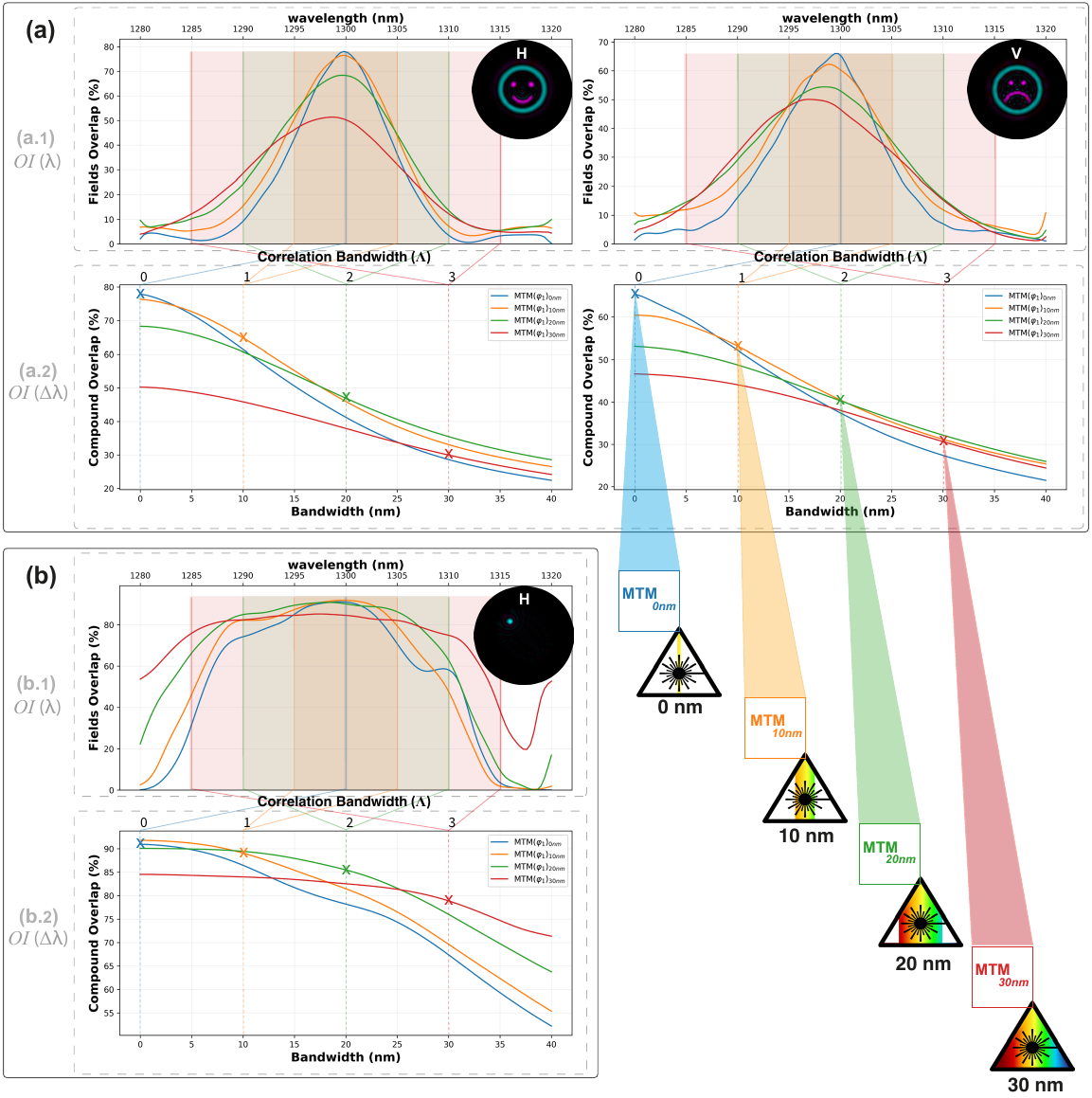}
    \caption{Study of the performance of each measured MTM$_{\Delta\lambda_S}$ for different SCS. Experimentally measured field overlap ($OI$) between the output fields and the target field as a function of $\lambda$ and $\Delta\lambda$ for, (a) smiley(H)-sad(V) face target pattern using most of the modes of the fibre simultaneously, (b) single-polarisation diffraction limited spot target pattern using just a few modes.}
    \label{fig:figure_short_results}
\end{figure*}

For the sake of simplicity and to perform only a single tomographic measurement, we used a tunable laser to emulate any desired SCS up to $\Delta\lambda_S = 40$~nm in post-processing as described in Methods~\ref{meth:SCS}. Knowing that our fibre-under-test exhibits a correlation bandwidth of $\Lambda \approx 10$~nm, as shown in Supplementary Note~\ref{sn:fibre-lambda}, we reconstructed four different tomographic measurements with emulated broadband sources at the input ($\Delta\lambda_S = 0$, 10, 20 and 30~nm), corresponding to up to 3$\Lambda$. The retrieved MTM$_{\Delta\lambda_S}$ associated with $\ket{\varphi_i}$ and its corresponding attenuation per channel plots are depicted in Fig.~\ref{fig:short_MTM_SVD}. One of each of these MTM$_{\Delta\lambda_S}$ maximises light control over all spatial-spectral channels for the given $\Delta\lambda_S$. As expected, as the $\Delta\lambda_S$ increases, the number of available singular values diminishes inside MTM($\varphi_1$), as establishing a phase relationship between all modes across all wavelengths gets more challenging. This effect becomes particularly significant when $\Delta\lambda_S\geq$ 3$\Lambda$. We explore the secondary eigenstates and their associated MTMs in Supplementary Note~\ref{sn:scs-extra}. In what follows, we focus on the strongest eigenstate for each emulated source bandwidth, which maximises light control with a single MTM. To assess the method's performance when an SCS is utilised, we generated distinct field patterns on the camera side, distinguishing between two scenarios, depending on the correlation bandwidth of each target ($\delta \lambda_T$): The first involves generating a pattern utilising the majority of the fibre's modes ($\delta\lambda_T \approx \Lambda$), encompassing both polarisation states simultaneously, as this requires completely arbitrary control of all spatial and polarisation modes. The second only involves a single polarisation and a limited number of spatial modes ($\delta\lambda_T > \Lambda$). These two scenarios' results are depicted in Fig.~\ref{fig:figure_short_results}, and represent the worst and best performance expectation cases, when using the retrieved MTM$_{\Delta\lambda_S}$. In both cases, we measure the overlap $OI{(\lambda)}$ between the ideal field pattern $\Vec{E}_{T}$ and the measured field $\Vec{E}_{M}(\lambda)$ at the output of the fibre as a function of wavelength for each MTM$_{\Delta\lambda_S}$,  Fig.~\ref{fig:figure_short_results}(a.1, b.1), with further details in Methods~\ref{meth:OI} and \ref{meth:pattern}. For a better interpretation of the results, we apply a wavelength-to-bandwidth transformation ($\lambda \rightarrow \Delta\lambda$), integrating the measured $OI{(\lambda)}$ across all available $\Delta\lambda$, Fig.~\ref{fig:figure_short_results}(a.2, b.2).

When trying to generate a smiley/sad face pattern ($\delta\lambda_T \approx \Lambda$) as a function of $\Delta\lambda$, as shown in Fig.~\ref{fig:figure_short_results}(a.2), the best $OI$ at a source-bandwidth $\Delta\lambda_S$ (marked with the colour-coded `$\times$' cross marks) this is achieved when employing the MTM$_{\Delta\lambda_S}$ for $\Delta\lambda \leq 2\Lambda$. This behaviour is intuitive since the number of valid $\ket{\varphi_i}$ increases with $\Delta\lambda_S$ promoting spatial decoherence. Hence, no well-defined phase relationship between some modes is found. In the attempt to control most of the modes across all spatial-spectral channels simultaneously, the generated pattern quality $OI$($\Delta\lambda$) will quickly drop. Crucially, note that the monochromatic (MTM$_{\Delta\lambda_S = 0\,\mathrm{nm}} =$ MTM$_{\lambda = 1300\,\mathrm{nm}}$) open channel solution (highest transmission eigenstate) does not provide the optimal broadband solution for $\Delta\lambda_S > 0$~nm, being always outperformed by another broadband MTM$_{\Delta\lambda_S}$. This confirms that SST provides the optimal solution for a given employed source. Another result that can be inferred from this analysis is the broadest-band MTM concept, corresponding to the transmission matrix which maximises the $\Delta\lambda$ operation range before -3~dB roll-off. In this scenario, for patterns $\delta\lambda_T \approx \Lambda$, it is achieved at $\Delta\lambda_S = 2\Lambda$ (green curve). To corroborate the consistency of the results, we repeated the analysis generating a random amplitude and phase superposition of all the modes in Supplementary Note~\ref{sn:scs-extra}. Besides following the same trend, we could identify for all generated patterns with MTM$_{2\Lambda}$ a consistent $\sqrt{2}$ bandwidth enhancement with respect to the monochromatic matrix, in such a way: $\delta\lambda'_T \approx \sqrt{2} \delta \lambda_T$, where $\delta\lambda '$ is the new correlation bandwidth of the target pattern generated with the MTM$_{2\Lambda}$.

Repeating the previous analysis for spot focusing ($\delta\lambda_T > \Lambda$), the overall trends are qualitatively similar, but performance is substantially improved. The operation range of all extracted MTMs is extended, with MTM$_{3\Lambda}$ offering the broadest range and generating spots with 80--85\% overlap across three times the fibre correlation bandwidth. This is expected because the spot involves only a few modes with similar group delays, reducing modal dispersion and preserving spatial coherence over a wider bandwidth. Nevertheless, we cannot verify that MTM$_{3\Lambda}$ is the true broadest-band MTM, nor transfer the $\sqrt{2}$ enhancement rule established for $\delta\lambda_T \approx \Lambda$ without further tests. Identifying the broadest-band MTM for spot focusing would require beam shaping over $\Delta\lambda \geq 3\delta\lambda_T$, where $\delta\lambda_T \approx 25$~nm for the focused spot ($\geq 75$~nm total), exceeding the 40~nm bandwidth emulated in our setup.

\section*{Discussion}

We showed that the MTM matrix of complex medium can be recovered from intensity-only measurements without an external reference, iterative phase retrieval, or prior selection of an internal reference channel. By applying spatial-state tomography, the complex modal relationships are reconstructed independently at each detector position, while the otherwise undetermined phase offsets between detection pixels are recovered from the spatial redundancy provided by oversampling the modal speckle grains. The method therefore reduces complex-field characterisation to a sequence of independent intensity measurements while retaining the phase information required for deterministic beam shaping for a wide range of light sources, from high to low coherent sources. Although a camera was used here to record all output positions in parallel, the measurement does not fundamentally require array detection: the same information could be acquired using a single photodiode combined with spatial scanning. This unlocks a series of new capabilities when using a broadband source at the input, for instance, the intensity measurements can be routed to a spectrometer to measure a spectrally resolved MTM($\lambda$) or to an oscilloscope to study temporal dynamics~\cite{ploschnerSpatialTomographyLight2022a}. The same principle could also be extended to self-referenced time-gated MTM measurements by replacing a conventional photodiode with photostable engineered nanoparticles exhibiting nonlinear multiphoton excitation~\cite{denkova_3d_2019}, together with a high-temporally coherent source. Because the nonlinear response is strongest when modal contributions arrive simultaneously, the detected signal would provide temporal discrimination between modal pathways and could enable reconstruction of a time-gated fibre response without a separate reference arm.

With respect to the necessary number of input fields to probe the fibre, without any prior knowledge of the fibre, the complete set of $2N^2-N$ analyser states is required to ensure that a reference is available for any and all modes in the fibre (or as few as $N^2$ in an ideal tomography measurement). Unlike existing self-referencing approaches, which involve launching uncontrolled random field patterns into the fibre with the hope of retrieving the fibre TM, SST ensures that a reliable reference will be found. While the number of analyser states scales unfavourably with the number of modes, making it potentially non-viable for certain applications, understanding the physics of the fibre (or the device under test) can help to mitigate this challenge without accuracy degradation. For example, in the case of GI MMFs, it is possible to reduce the number of projections by performing small tomographic measurements targeting individual mode groups ($n_g$), knowing that modes between mode groups do not couple. Applying this approach, the analyser states scales as $\sum_g \mathcal{O}(n_g^2)$ rather than $\mathcal{O}((\sum_g n_g)^2)$, omitting inter-group pairwise projections that target matrix elements absent from the recovered MTM. Moreover, projection hardware acceleration via a DMD or PLM (Phase-only Light Modulator)\cite{rochaFastLightefficientWavefront2024} together with mode-group partitioning could bring per-wavelength SST acquisition to tens of seconds, enabling either sequential tunable-laser measurements of a self-referenced  MTM($\lambda$), or sparse-wavelength SST combined with parametric dispersion modelling~\cite{lee_efficient_2023} to recover the full spectral transfer matrix efficiently.

We demostrated that SST performs at the level of off-axis digital holography (gold standard) when using an LCS. This could make our proposed self-reference technique attractive even when a reference is available. Nevertheless, the true power of the proposed approach emerges when operating with broadband sources. In this scenario, each wavelength will experience a different MTM($\lambda$) and consequently there is not a unique solution that can control all the light. We experimentally demonstrated that SST can effectively retrieve the MTM$_{\Delta\lambda_S}$ associated with the dominant eigenstate ($\varphi_1$), which optimises the light control for that particular source. Furthermore, we demonstrated source-matched beam shaping at the distal fibre facet: MTM$_{\Delta\lambda_S}$ provides the optimal solution for a source of bandwidth $\Delta\lambda_S$, as confirmed across target patterns involving different numbers of modes (Fig.~\ref{fig:figure_short_results}). Under these conditions there is no straightforward interferometric benchmark, or gold standard, against which to compare. Comparable optimisation could, in principle, be achieved by measuring a spectrally resolved MTM($\lambda$) with an external reference and a swept laser, then constructing an operator that weights the monochromatic matrices according to the source spectrum to derive an equivalent MTM$_{\Delta\lambda_S}$. We hypothesise that the resulting operator is closely related to the broadband flux matrix $\mathbf{A} = \int \mathbb{S}(\lambda)\, t(\lambda)^{\dagger} t(\lambda)\, d\lambda$ described in~\cite{hsu_broadband_2015}, where $t(\lambda)$ is the monochromatic transmission matrix and $\mathbb{S}(\lambda)$ the spectral density. However, suitable swept or tunable sources may not exist over the required spectral range. SST instead performs this optimisation directly with the intended source, automatically incorporating its bandwidth and coherence properties.
 
We are convinced that the vast amount of information that SST delivers can be further exploited when combined with SCS. In our current study, we have focused mostly on the highest eigenstate ($\varphi_1$), but it is important to note that the remaining eigenstates with associated MTM($\varphi$) hold valuable insights for controlling the remaining light. MTM($\varphi_1$) ensures that on average across all modes and wavelengths, the control of light is maximal. However, it may not guarantee optimal performance for a specific target pattern. For instance, we previously showed that certain modes/mode groups in MTM$_{3\Lambda}$ are not part of this solution as they lack any phase relationship with the rest of the modes. Then, in certain cases where patterns are predominately composed of those ``missing" modes, alternative MTM($\varphi_i$) may be more effective. Similarly to other wavefront shaping techniques based on the evaluation of differential changes between multiple matrices $t(\alpha)$ along an arbitrary property $\alpha$---such as the generalised Wigner-Smith operator $Q_{\alpha} = -jt^{-1}\partial_{\alpha}t$~\cite{ambichl_focusing_2017, matthes_learning_2021}, the Fisher information $F_{\alpha} = (\partial_{\alpha} t)^{\dagger} t \partial_{\alpha}t$ and discrimination operator~\cite{bouchet_maximum_2021, bouchet_optimal_2021}---it could be possible considering all MTM($\varphi$) to generate a transformation operator to perform an incoherent superposition of modes to enhance specific targets. This operator would be target-dependent, allowing tailored transformations to suit the desired target pattern at the distal end of the fibre across a specific $\Delta\lambda$. This opens up new possibilities for enhancing beam shaping through scattering media when working with broadband sources using a single-phase mask at the input.

\section*{Author contributions}
MMM built the optical apparatus and carried out the experiments with advice from JC, MP, MM, and AK. Code for experimental automation and Stokes data analysis was authored by MMM with input from MP and JC. MMM led the data analysis with contributions from JC, MP, AK, and MM. The manuscript was drafted by MMM, with contributions from MP and JC and critical feedback from AK and MM. JC and MP co-supervised the project, assisted by MM. Conceptualisation was led by JC, with contributions from MMM, MP, and MM.

\section*{Funding and Acknowledgements}
This research was supported by an Australian Government Research Training Program (RTP) Scholarship and by the Australian Research Council (ARC) funding through ARC Discovery Project (DP250101269), ARC Future (FT230100388 and FT220100103), ARC DECRA (DE210100934) and Westpac Scholars Trust.

During the preparation of this work, the author(s) used ChatGPT (OpenAI), Claude (Anthropic), and Gemini (Google) to assist with language editing and improve the clarity of the manuscript. Cursor (Anysphere) was used to assist with \LaTeX{} formatting and typesetting. After using these tools, the author(s) reviewed and edited the content as needed and take full responsibility for the content of this publication.

\section*{Disclosures and competing interests}
The authors declare no competing interests.

\section*{Data availability}
The data that support the findings of this study are available from the corresponding author upon reasonable request.
Processing code and example tomographic measurement data (Version~0.1.0, arXiv snapshot) are available at
\url{https://github.com/MarKo7s/SST_MMF_characterization}
and archived on Zenodo under DOI~\href{https://doi.org/10.5281/zenodo.21637820}{10.5281/zenodo.21637820}.
\begin{center}
\href{https://github.com/MarKo7s/SST_MMF_characterization}{\includegraphics[height=11pt]{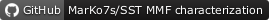}}\hspace{0.4em}%
\href{https://github.com/MarKo7s/SST_MMF_characterization/releases/tag/v0.1.0}{\includegraphics[height=11pt]{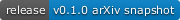}}\hspace{0.4em}%
\href{https://doi.org/10.5281/zenodo.21637820}{\includegraphics[height=11pt]{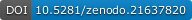}}
\end{center}

%\section*{REFERENCES}
%\nocite{*}

\section*{Methods}

\subsection{Phase locking between camera pixels}
\label{meth:phase}

Each eigenstate at a given pixel is expressed as a modal decomposition in an orthogonal basis of choice $\{\ket{\psi_n}\}$ spanning the $N$-dimensional fibre mode space:
\begin{equation}
    \ket{\varphi} = \sum_{n=0}^{N-1} c_n \ket{\psi_n},
    \label{eq:eigenstate_decomposition}
\end{equation}
where $c_j$ are complex coefficients, such that any pure state is described as a vector $(c_0, c_1, \ldots, c_{N-1})$ in the chosen basis. The phase-locking procedure can be performed in the vector space or on the equivalent reconstructed field.

To establish the phase relationship between all camera pixels, the global phase between spots within the same speckle grain is recovered from the inner product between a phase-locked reference pixel and an unlocked target pixel, yielding the phase offset and corrected target eigenstate:
\begin{align}
    \Delta\phi_{\mathrm{ref|target}}
    &= \arg\!\left(
    \langle \varphi'_{\mathrm{ref}} | \varphi_{\mathrm{target}} \rangle
    \right)
    \label{eq:phase_offset} \\
    &= \arg\!\left(
    \sum_{n=0}^{N-1} \varphi'^{*}_{\mathrm{ref},n}\,\varphi_{\mathrm{target},n}
    \right)
    \notag \\
    \varphi'_{\mathrm{target}}
    &= \varphi_{\mathrm{target}}\, e^{-j\Delta\phi_{\mathrm{ref|target}}}.
    \label{eq:phase_correct}
\end{align}

A general logic to follow to lock all the pixels is shown in Fig.~\ref{fig:phase_locking_flow}.

\begin{center}
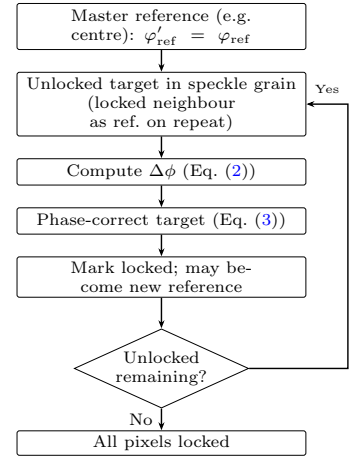

    \scalebox{0.90}{%
    \begin{tikzpicture}[
        node distance=0.32cm,
        box/.style={draw, rectangle, rounded corners=1pt, align=center,
                    text width=4.1cm, inner sep=2pt, font=\scriptsize},
        decision/.style={draw, diamond, aspect=2.2, align=center,
                         inner sep=0.5pt, font=\scriptsize},
        arrow/.style={-{Stealth[length=1.2mm]}, semithick}
    ]
        \node[box] (b1) {Master reference (e.g.\ centre): $\varphi'_{\mathrm{ref}}=\varphi_{\mathrm{ref}}$};
        \node[box, below=of b1] (b2) {Unlocked target in speckle grain\\(locked neighbour as ref.\ on repeat)};
        \node[box, below=of b2] (b3) {Compute $\Delta\phi$ (Eq.~\eqref{eq:phase_offset})};
        \node[box, below=of b3] (b4) {Phase-correct target (Eq.~\eqref{eq:phase_correct})};
        \node[box, below=of b4] (b5) {Mark locked; may become new reference};
        \node[decision, below=0.45cm of b5] (check) {Unlocked\\remaining?};
        \node[box, below=of check] (done) {All pixels locked};

        \draw[arrow] (b1) -- (b2);
        \draw[arrow] (b2) -- (b3);
        \draw[arrow] (b3) -- (b4);
        \draw[arrow] (b4) -- (b5);
        \draw[arrow] (b5) -- (check);
        \draw[arrow] (check.south) -- node[left, font=\scriptsize] {No} (done.north);
        \draw[arrow] (check.east) -- ++(1.45,0) |- node[near end, above, font=\tiny, yshift=1pt] {Yes} (b2.east);
    \end{tikzpicture}%
    }
    \captionof{figure}{Iterative phase-locking logic. A master reference pixel seeds the global phase; each unlocked pixel is corrected relative to a locked neighbour within the same speckle grain until all pixels are locked.}
    \label{fig:phase_locking_flow}
\end{center}

\subsection{Beam shaping quality: Overlap integral}
\label{meth:OI}

For beam quality analysis we employ the wavelength-dependent overlap integral $OI(\lambda)$, Eq.~\ref{eq:OI}. 

\begin{equation}
    OI(\lambda) = \frac{\left| \iint \limits_S \Vec{E}_{1}(\lambda) \cdot  \Vec{E}_{2}^{*}(\lambda) \, dS \right|^2}{\iint \limits_S |\Vec{E}_{1}(\lambda)|^2 \, dS \iint \limits_S  |\Vec{E}_{2}(\lambda)|^2 \, dS}
    \label{eq:OI}
\end{equation}

The $OI(\lambda)$ measures the complex similarity between the target field $\Vec{E}_{1}(\lambda)$ and the experimentally measured field $\Vec{E}_{2}(\lambda)$ at wavelength $\lambda$, with values between 0 and 1. Both fields are acquired by off-axis digital holography at each sampled wavelength. Evaluating $OI(\lambda)$ rather than intensity alone presents a more stringent criterion: for instance, if we have no control over any fibre mode, the intensity overlap between a random speckle pattern ($\Vec{E}_{2}(\lambda)$) and a smiley-face target ($\Vec{E}_{1}(\lambda)$) is unlikely ever to be zero, which can misleadingly suggest residual control. For narrowband (LCS) measurements at a single wavelength $\lambda_c$, Eq.~\ref{eq:OI} reduces to one evaluation, $OI(\lambda_c)$. For broadband (SCS) characterisation, the same spatial target pattern is maintained across wavelengths ($\Vec{E}_{1}(\lambda) \equiv \Vec{E}_{T}(\lambda)$) and $OI(\lambda)$ is computed at each $\lambda$ by overlapping the holographically retrieved measured field $\Vec{E}_{2}(\lambda) \equiv \Vec{E}_{M}(\lambda)$ with $\Vec{E}_{1}(\lambda)$.

\subsection{Fibre optic singular values}
\label{meth:SVD}

For an ideal mode transmission matrix describing the propagation of light in an optical fibre, the singular values would be expected to be relatively flat, and near 0~dB for modes far from the cut-off, where the cutoff corresponds to higher-order modes that are not fully supported by the fibre. How close the measured singular values are to 0~dB for modes far from cutoff is an indicator of the measurement accuracy, as the fibre itself should have little to no mode-dependent loss for such modes, and any gradual attenuation roll-off before cut-off is more likely to reflect mode-dependent loss in the mode excitation itself than in the propagation along the length. As a practical figure of merit, we establish a -3~dB cutoff threshold: channels beyond this point are considered too attenuated to be usable. Compared with off-axis digital holography, SST yields not only more channels above this threshold, but also singular values closer to the ideal 0~dB level, consistent with a more physically realistic description of the fibre. The presence of artefacts in the measurement will directly translate into a reduction of the number of singular values close to 0~dB since they will break the orthogonality of the matrix.

\subsection{Short coherence length source emulation}
\label{meth:SCS}

For each of the $2N^2-N$ input analyser states excited at the proximal end of the fibre, $D$ equally spaced optical-frequency ($\delta\nu$) frames are recorded using a swept-wavelength interferometer~\cite{MaestreMorote2026pySSTri}. Because the camera is much faster than the SLM, the tomographic measurement duration has negligible relationship with the number of optical frequencies ($\nu_i$) triggered. In this experiment the laser is swept from 1280~nm ($\lambda_{\min}$) to 1320~nm ($\lambda_{\max}$), corresponding to a maximum bandwidth of 40~nm ($\Delta\nu \approx 7.1$ THz), and $D=118$ frames spaced by $\delta\nu \approx 60.2 $ GHz. The $I(x,y,k)_{\lambda}$ are acquired for every analyser state~$k$ forming a raw archive equivalent to $D$ independent monochromatic measurements (one per triggered $\lambda \equiv \nu$). An SCS is emulated by adding incoherently the frame intensities along wavelength axis for each~$k$. Finally, an arbitrary source of bandwidth $\Delta\lambda_S=\lambda_2-\lambda_1$ is synthesised in post-processing as in Eq.~\ref{eq:emulate_SCS}.

\begin{equation}
    I_{\Delta\lambda_S}(x,y,k) = \sum_{i = \lambda_1}^{\lambda_2} I(x,y,k)_i \label{eq:emulate_SCS}
\end{equation}

Note that the optical-frequency sampling interval must satisfy the
Nyquist-type condition $\delta\nu\le\Lambda/2$ to resolve the fibre
spectral structure (here $\delta\nu\approx60.2$~GHz and
$\Lambda\approx1.8$~THz). In practice, $D$ is limited by available
memory unless the incoherent sum in Eq.~\eqref{eq:emulate_SCS} is
performed on the fly, so that individual wavelength frames need not be
retained for post-processing. Further details of the full setup, including the camera trigger, are given in Supplementary Note~\ref{sn:lab-setup}.

\subsection{Pattern generation}
\label{meth:pattern}

The generation of any desired pattern at the end of the fibre is done by propagating the fields through the measured MTM. Firstly, any desired target fields ($\Vec{E}_{T}$) are decomposed in the mode basis describing the MTM, in this case, LG-basis. The needed input fields ($\Vec{E}_{IN}$) to generate the desired $\Vec{E}_{T}$ are calculated by reconstructing the back-propagated mode coefficients through the fibre. The field is experimentally generated by the same GS algorithm used for the SST analyser state fields. The resulting fields at the output ($\Vec{E}_{OUT}$) are measured using an external reference and off-axis digital holography, also employed to measure MTM$_{DH}$. The theoretical best patterns ($\Vec{E}_{B}$) achievable with a measured MTM are obtained by decomposing the target $\Vec{E}_{T}$ in the same mode basis, back-propagating its coefficients through the fibre, and forward-propagating them again. This round trip defines the optimal output implied by the matrix itself, so any attenuated or weakly transmitted channels encoded in the measured singular values are already included. For an ideal lossless fibre the MTM is unitary, so back-propagation is equivalent to applying the conjugate transpose ($\mathbf{M}^{-1}=\mathbf{M}^{\dagger}$) and $\Vec{E}_{B}$ matches $\Vec{E}_{T}$. Departures from unitarity in the measured MTM drops round trip quality in direct correspondence with the singular-values, providing an upper bound on pattern quality.
\ifdefined\paperloadedbymain
\else
\end{document}
\fi

% --- Supplementary Information ---
\clearpage
\onecolumngrid
% Notes/subheads left-aligned; SI title block is centered separately.
\makeatletter
\renewcommand{\subsection}{\@startsection{subsection}{2}{\z@}%
  {-3.25ex\@plus -1ex \@minus -.2ex}{1.5ex \@plus .2ex}%
  {\normalfont\large\bfseries\raggedright}}
\renewcommand{\subsubsection}{\@startsection{subsubsection}{3}{\z@}%
  {-3.25ex\@plus -1ex \@minus -.2ex}{1.5ex \@plus .2ex}%
  {\normalfont\normalsize\bfseries\raggedright}}
\makeatother
\setcounter{supplnote}{-1}% first note is 0
\setcounter{figure}{0}
\setcounter{table}{0}
\setcounter{equation}{0}
\renewcommand{\thefigure}{S\arabic{figure}}
\renewcommand{\thetable}{S\arabic{table}}
\renewcommand{\theequation}{S\arabic{equation}}
\begin{center}
  {\normalsize\bfseries
    Supplementary Information\\[0.25em]
    Broadband Control of Light through Complex Media via Automatic
    Self-Referencing Transmission Matrix Characterisation}\\[1em]
  {\small
    M.~Maestre Morote$^{1,\ast}$,
    M.~Mounaix$^{1}$,
    A.~Komonen$^{1}$,
    M.~Pl\"{o}schner$^{1}$,
    J.~Carpenter$^{1}$}\\[0.5em]
  {\footnotesize\itshape
    $^{1}$School of Electrical Engineering and Computer Science,
    The University of Queensland, Brisbane, QLD 4072, Australia.}\\[0.3em]
  {\footnotesize $^{\ast}$Corresponding author: m.maestremorote@uq.edu.au}
\end{center}
\vspace{1em}
\newcommand{\siloadedbymain}{}
% Typeset SI bibliography once: SI uses si_* cite keys (sup.bib / supplementary.bbl)
% so numbers [1]… never collide with main.bbl / paper.bib.
\setbox\sibibbox=\vbox{%
  \hsize=\textwidth
  \linewidth=\hsize
}
% !TEX root = supplementary.tex
\subsection*{Supplementary video:\href{https://www.youtube.com/watch?v=0rNlo3MXHnQ&ab_channel=MarcosMaestreMorote}{YouTube link}}

\begin{enumerate}
    \item {\href{https://www.youtube.com/watch?v=0rNlo3MXHnQ&ab_channel=MarcosMaestreMorote}{Motivation} - {0s}}
    \item {\href{https://youtu.be/0rNlo3MXHnQ?si=VGOJYZ0DMMFNuAlD&t=93}{Multimode fibre characterisation} - {1m:33s}}
    \item {\href{https://youtu.be/0rNlo3MXHnQ?si=x1UeEI4UQMDYuksd&t=168}{Stokes polarimetry analogy} - {2m:48s}}
    \item {\href{https://youtu.be/0rNlo3MXHnQ?si=YdTXIV1tGGJT_8RJ&t=360}{Spatial state tomography} - {6m:0s}}
    \item {\href{https://youtu.be/0rNlo3MXHnQ?si=LrbFosJCik3ATqML&t=505}{Mode transmission matrix retrieval} - {8:25s}}
    \item {\href{https://youtu.be/0rNlo3MXHnQ?si=yYxVnz6vXFTirk-P&t=591}{Long coherence source results} - {9m:50s}}
    \item {\href{https://youtu.be/0rNlo3MXHnQ?si=KtShUJyK-Q3-fBdp&t=741}{Short coherence source results} - {12m:21s}}
    \item {\href{https://youtu.be/0rNlo3MXHnQ?si=ggEk9F-AFGQ8bzDW&t=957}{Conclusions} - {15m:57s}}
\end{enumerate}

\subsection*{Supplementary code:
\href{https://github.com/MarKo7s/SST_MMF_characterization}{\includegraphics[height=11pt]{figures/badges/github.png}}\hspace{0.4em}%
\href{https://github.com/MarKo7s/SST_MMF_characterization/releases/tag/v0.1.0}{\includegraphics[height=11pt]{figures/badges/release.png}}\hspace{0.4em}%
\href{https://doi.org/10.5281/zenodo.21637820}{\includegraphics[height=11pt]{figures/badges/zenodo_doi.png}}}

Processing routines and data to retrieve the MTM of a simulated GI fibre and of the experimental OM1 fibre (five mode groups, five $\Delta\lambda_S$), Version~0.1.0 (arXiv snapshot); DOI~\href{https://doi.org/10.5281/zenodo.21637820}{10.5281/zenodo.21637820}.
Notebooks: \href{https://github.com/MarKo7s/SST_MMF_characterization/blob/main/SST_MTM_retrival_example_simulation.ipynb}{simulation}, \href{https://github.com/MarKo7s/SST_MMF_characterization/blob/main/SST_MTM_retrival_example_experimental.ipynb}{experimental}.

\subsection*{Scope}

Evaluation metrics (overlap integral $OI$, singular-value analysis, pattern generation, and SCS emulation) and phase locking between camera pixels are defined in the main article Methods. This Supplementary Information provides:
\begin{itemize}
    \item Note~\ref{sn:sst-theory} --- SST theory;
    \item Note~\ref{sn:mmf-mtm} --- MMF characterisation and MTM assembly;
    \item Note~\ref{sn:lab-setup} --- extended laboratory setup, alignment, and data handling;
    \item Note~\ref{sn:lcs-extra} --- additional LCS figures with interpretation;
    \item Note~\ref{sn:fibre-lambda} --- fibre correlation bandwidth $\Lambda$;
    \item Note~\ref{sn:scs-extra} --- secondary eigenstates and extended SCS beam shaping.
\end{itemize}

\supplnote{sn:sst-theory}{Spatial state tomography background}

We provide a brief overview of state tomography using quantum formalism before plunging into the parallelised high-dimensional spatial tomography used to characterise multimode fibre. Spatial state tomography (SST) is a generalisation of state tomography applied to the spatial degree of freedom of light. It can be visualised as the high-dimensional counterpart to Stokes polarimetry.

In this section we index fibre modes by $n\in\{0,\ldots,N-1\}$ in the basis $\{\ket{\psi_n}\}$; density-matrix eigenstates $\ket{\varphi_i}$ by $i\in\{1,\ldots,N\}$; Gell--Mann matrices and Stokes parameters by $j\in\{0,\ldots,N^2-1\}$ ($\hat{\sigma}_j$, $S_j$); the unique eigenstates of each $\hat{\sigma}_j$ by $r$ ($\ket{\Omega_j^{r}}$, $\kappa_j^r$), when referenced to a particular $j$-matrix or in a flat indexing format by $k\in\{1,\ldots,m\}$ with $m=2N^2-N$.

\subsubsection*{Density-matrix formalism}

An unknown state of light, which could be a pure state or a mixed state (i.e., an incoherent superposition of multiple pure states), is described by a density matrix $\hat{\rho}$. By the spectral theorem, it can be represented as its eigendecomposition:
\begin{equation}
    \hat{\rho} = \sum_{i=1}^{N} p_i\ket{\varphi_i}\bra{\varphi_i},
\end{equation}
where $p_i$ is the probability (or intensity weighting) of finding a given $\ket{\varphi_i}$. The pure eigenstates $\ket{\varphi}$ in the mixture $\hat{\rho}$ are linear combinations of the $N$ modes forming the basis $\{\ket{\psi_n}\}$ along the degree of freedom in which the tomography is performed:
\begin{equation}
    \ket{\varphi} = \sum_{n=0}^{N-1} c_n \ket{\psi_n},
    \label{eq:eigenstate_reconstruction}
\end{equation}
where $c_n$ are complex coefficients.

To reconstruct $\hat{\rho}$, a tomographically complete set of observables must be measured. For an $N$-level system these are the generalised Gell--Mann matrices $\hat{\sigma}_j$: $N^2-1$ linearly independent $N\times N$ traceless Hermitian matrices that, together with the identity, span the space of density matrices. With normalisation $\Tr(\hat{\sigma}_j\hat{\sigma}_\ell)=2\delta_{j\ell}$, the density matrix is recovered as a weighted sum:
\begin{equation}
    \hat{\rho} = \frac{1}{2} \sum_{j=0}^{N^2-1} S_j \hat{\sigma}_j,
    \label{eq:rho_recon_SI}
\end{equation}
where $S_j=\Tr(\hat{\rho}\,\hat{\sigma}_j)$ are the Stokes parameters collected in the high-dimensional Stokes vector $\mathbf{S}$. How these expectation values are obtained experimentally is developed below.

\subsubsection*{Observable expectation, Gell-mann matrices, projection operator and Stokes vector}

Projective measurements are the specific outcomes used to infer physical properties of an unknown system; in the tomographic formalism these quantities are called observables. An observable is a linear Hermitian operator $\hat{O}$ acting on the $N$-dimensional Hilbert space. For a mixed state $\hat{\rho}$, the expectation value follows the Born rule:
\begin{equation}
    \langle O \rangle_{\hat{\rho}}
    = \sum_{i=1}^{N} p_i\bra{\varphi_i}\hat{O}\ket{\varphi_i}
    = \Tr(\hat{\rho}\,\hat{O}),
    \label{eq:observable_expectation_SI}
\end{equation}
where $\Tr$ denotes the trace. The Gell--Mann matrices $\hat{\sigma}_j$ generalise the Pauli matrices from SU(2) to SU($N$), forming an orthogonal basis for the Lie algebra of traceless Hermitian operators.

The expectation values $S_j$ of finding an observable $\hat{\sigma}$ in the mixture is not always directly measurable. Instead each $\hat{\sigma}_j$ is spectrally decomposed into projection operators onto its eigenstates, the analyser states $\ket{\Omega_j^{r}}$:
\begin{equation}
    \hat{\sigma}_j = \sum_r \kappa_j^r \ket{\Omega_j^{r}}\bra{\Omega_j^{r}}
    = \sum_r \kappa_j^r \hat{P}_j^r,
    \label{eq:gellmann_spectral_SI}
\end{equation}
where $\kappa_j^r$ are the corresponding eigenvalues and $\hat{P}_j^r=\ket{\Omega_j^{r}}\bra{\Omega_j^{r}}$ the projection operator for the eigenvectors. Projecting each analyser state onto $\hat{\rho}$ yields the Stokes component:
\begin{equation}
    S_j = \Tr(\hat{\rho}\,\hat{\sigma}_j)
    = \sum_r \kappa_j^r \bra{\Omega_j^{r}}\hat{\rho}\ket{\Omega_j^{r}}.
    \label{eq:expectation_projections_SI}
\end{equation}
Similarly, if an observable comprises several eigenstates, each must be projected individually to remove ambiguity. One of each of the Gell--Mann matrices therefore explores a subspace of the $N\times N$ Hilbert space and projecting its eigenstates one-by-one provides unambiguous intensity outcomes. The detected intensity $I_{\ket{\Omega_k}}=\bra{\Omega_k}\hat{\rho}\ket{\Omega_k}$ per analiser state is equivalent to the integral overlap over the detection area ($A$) between all the fields inside the mixture ($\varphi_i(A)$) and the projected field ($\Omega_k(A)$):

\begin{equation}
    I_{\ket{\Omega_k}} = \braket{\Omega_k | \hat{\rho} | \Omega_k}
    = \sum_{i} p_i \braket{\Omega_k | \varphi_i}\braket{\varphi_i | \Omega_k}
    = \sum_i p_i \left| \braket{\Omega_k | \varphi_i} \right|^2
    = \sum_i p_i \left| \int \Omega_k^{\ast}(A)\, \varphi_i(A)\, dA \right|^2
    \label{eq:intensity_probability}
\end{equation}

Finally another way to visualise the Stokes vector $\mathbf{S}$ is as the Cartesian coordinates of $\hat{\rho}$ in the Gell--Mann basis (Eq.~\eqref{eq:rho_recon_SI}), living in a real hypersphere of dimension $N^2-1$. Purity sets the radius: the maximally mixed state is at the centre ($\mathbf{S}=0$), pure states lie on the surface, and partially mixed states in between. For $N=2$ this is exactly the Poincar\'{e} sphere, with $(S_1,S_2,S_3)$ as the familiar polarisation Stokes parameters (following subsection).

\subsubsection*{Tomographic setup and efficient reconstruction}

The physical measurement requires projecting a total of $m=2N^2-N$ unique analyser states $\ket{\Omega_k}$ onto the unknown state. These are the eigenvectors of $\hat{\sigma}_j$, selected with a uniqueness threshold $\Theta=1-1/N^2$: a new eigenstate $\ket{\Omega_j^{r}}$ is retained as $\ket{\Omega_k}$ only if it is sufficiently distinct from all previously stored states ($\bra{\Omega_k}\Omega_j^{r}\rangle<\Theta$). The associated eigenvalues $\kappa_j^r$ populate a sparse weighting matrix $\mathbf{K}$ of dimension $N^2\times(2N^2-N)$, with $\mathbf{K}_{jk}=\kappa_j^r$ when analyser state $k$ corresponds to the $r$-th eigenstate of $\hat{\sigma}_j$, and zero otherwise.

After acquiring the intensity of each projection, the Stokes vector is recovered by matrix multiplication:
\begin{equation}
    \mathbf{S} =\mathbf{K}\,\cdot \mathbf{I}_\Omega,
    \label{eq:stokes_matrix_SI}
\end{equation}
where $\mathbf{I}_\Omega(x,y)$ is the intensity vector with entries $I_{\ket{\Omega_k}} = \bra{\Omega_k}\hat{\rho}\ket{\Omega_k}$. Formulating the problem with help of Eq.~\eqref{eq:stokes_matrix_SI} allows to extend $\mathbf{S}$ to parallel SST problems by stacking independent $\mathbf{I}_\Omega$ vectors columns and performing the matrix multiplication.

\subsubsection*{Non-ideal tomography}

The framework above assumes ideal conditions: a perfectly matched basis, projections executed with $100\%$ accuracy, noiseless detection, and a static system throughout the acquisition. In practice, imperfections introduce errors that fall into three categories: (1) errors in the measurement basis, (2) counting and statistical errors, and (3) experimental-stability errors such as source-intensity, mechanical, or thermal drift. Any of these perturb the measured expectation values, so the retrieved state always deviates somewhat from the true one; in the Stokes-polarimetry picture the solution is no longer a single point on the Poincar\'{e} sphere but lies within a error uncertainty ``cloud of points''.

A rigorous error analysis is non-trivial. Conventionally it relies on repeating the tomography many times and fitting each error source to its statistical distribution via maximum-likelihood estimation (MLE) or Bayesian methods, which both requires a model per error source and scales poorly, as the number of projections grows rapidly with the system dimension. Even then, the recovered matrix is not guaranteed to be physical, so a final projection onto the nearest positive semi-definite (PSD) matrix is required \cite{si_higham_computing_1988}. For our system this counting-statistics approach is impractical: the dimensionality is simply too large to repeat the measurement.

Instead, we shrink the ``error uncertainty cloud'' proactively through calibration---aberration correction, mask efficiency optimisation, and frame averaging where possible---and then compute the nearest PSD density matrix and threshold sub-noise eigenvalues before extracting the eigenstates. Some errors, notably mechanical and thermal drift over extended acquisitions, remain beyond calibration: when the system drifts faster than it can be measured, previously inactive states gain probability weight and the mixture artificially grows, an effect that becomes more pronounced at higher dimension. Crucially, this noise predominantly corrupts the eigenvalues, which are largely irrelevant to our application, while the eigenvectors (the spatial patterns) remain robust. The measurement can be viewed as a power-maximisation problem: after probing with all projections, the primary eigenstate emerges as the linear superposition of basis modes that maximises transmitted power to that output channel. Drift and noise redistribute probability into additional, weakly weighted eigenstates, so the weight ($p_1$) of the physical solution may decrease; the modal superposition itself, however, remains essentially unchanged. That eigenvalue is only a predicted power fraction for that superposition. Because we seek the optimal focusing solution rather than an accurate absolute power forecast, this reduction in $p_1$ does not prevent reliable recovery of the principal eigenstate.

For these reasons we avoid measurement-reduction strategies---such as projecting only $N^2$ states and inferring the remainder, or compressed sensing with fewer than $N^2$ projections---which could shorten acquisition but introduce additional errors from power fluctuations, projection imperfections, or non-convex optimisation. We therefore project all $m=2N^2-N$ analyser states, accepting longer acquisition in exchange for tomographic completeness.

\subsubsection*{Stokes polarimetry analogy ($N=2$)}

For $N=2$, SST reduces to standard Stokes polarimetry: the Gell--Mann observables are the Pauli matrices,
\begin{align}
    \hat{\sigma}_0&= \begin{pmatrix}
    1 & 0 \\
    0 & 1
    \end{pmatrix}
    &
    \hat{\sigma}_1&=\begin{pmatrix}
    0 & 1 \\
    1 & 0
    \end{pmatrix}
    &
    \hat{\sigma}_2&=\begin{pmatrix}
    0 & -i \\
    i & 0
    \end{pmatrix}
    &
    \hat{\sigma}_3&= \begin{pmatrix}
    1 & 0 \\
    0 & -1
    \end{pmatrix}
    \label{eq:Pauli_matrices_SI}
\end{align}
and $\mathbf{S}=(S_0,S_1,S_2,S_3)$ gives Cartesian coordinates on the Poincar\'{e} sphere (Eq.~\eqref{eq:rho_recon_SI}). Each $\hat{\sigma}_j$ decomposes as in Eq.~\eqref{eq:gellmann_spectral_SI}; for $N=2$ the $m=6$ analyser states are the familiar polarisation bases ($\ket{H},\ket{V},\ket{D},\ket{A},\ket{R},\ket{L}$), with orthogonal eigenpairs on opposite Poincar\'{e}-sphere axes. Table~\ref{tab:stokes_polarimetry_setup} lists eigenvalues, eigenvectors, analyser states, and the Stokes components from Eq.~\eqref{eq:expectation_projections_SI}.

\begingroup
\setlength{\tabcolsep}{8pt}
\renewcommand{\arraystretch}{1.5}

\begin{table}[htbp]
\begin{tabular}{cccccccc}
\textbf{Pauli matrix} &
  \multicolumn{2}{c}{\textbf{eigenvalues}} &
  \multicolumn{2}{c}{\textbf{eigenvectors}} &
  \multicolumn{2}{c}{\textbf{Analyser state}} &
  \textbf{Stokes component} \\
\multicolumn{1}{c}{from Eq.~\eqref{eq:Pauli_matrices_SI}} &
  \multicolumn{4}{c}{as $\hat{\sigma}_j = U\cdot Y \cdot U^{-1}$} &
  \multicolumn{2}{c}{applying Eq.~\eqref{eq:eigenstate_reconstruction}} &
  \multicolumn{1}{c}{applying Eq.~\eqref{eq:expectation_projections_SI}} \\
 &
  $\kappa_j^0$ &
  \multicolumn{1}{c}{$\kappa_j^1$} &
  $v_j^0$ &
  \multicolumn{1}{c}{$v_j^1$} &
  $\ket{\Omega_j^0}$ &
  \multicolumn{1}{c}{$\ket{\Omega_j^1}$} &
  $S_j$ \\
\multicolumn{1}{l}{} &
  \multicolumn{1}{l}{} &
  \multicolumn{1}{l}{} &
  \multicolumn{1}{l}{} &
  \multicolumn{1}{l}{} \\
$\hat{\sigma}_0$ &
  $1$ &
  \multicolumn{1}{c}{$1$} &
  \begin{tabular}[c]{@{}c@{}}$\begin{pmatrix}     1  \\     0    \end{pmatrix}  $\end{tabular} &
  \multicolumn{1}{c}{\begin{tabular}[c]{@{}c@{}}$\begin{pmatrix}     0  \\     1     \end{pmatrix}  $\end{tabular}} &
   $\ket{H}$ &
  \multicolumn{1}{c}{$\ket{V}$}  & $\bra{H}\hat{\rho}\ket{H} + \bra{V}\hat{\rho}\ket{V}$
   \\
\multicolumn{1}{l}{} &
  \multicolumn{1}{l}{} &
  \multicolumn{1}{l}{} &
  \multicolumn{1}{l}{} &
  \multicolumn{1}{l}{} \\
$\hat{\sigma}_1$ &
  $1$ &
  \multicolumn{1}{c}{$-1$} &
  \begin{tabular}[c]{@{}c@{}}$\begin{pmatrix}     \frac{1}{\sqrt{2}}  \\     \frac{1}{\sqrt{2}}     \end{pmatrix}  $\end{tabular} &
  \multicolumn{1}{c}{\begin{tabular}[c]{@{}c@{}}$\begin{pmatrix}     \frac{1}{\sqrt{2}}  \\     \frac{-1}{\sqrt{2}}     \end{pmatrix}  $\end{tabular}} &
   $\ket{D}$&
  \multicolumn{1}{c}{$\ket{A}$} & $\bra{D}\hat{\rho}\ket{D} - \bra{A}\hat{\rho}\ket{A}$
   \\
  \multicolumn{1}{l}{} &
  \multicolumn{1}{l}{} &
  \multicolumn{1}{l}{} &
  \multicolumn{1}{l}{} &
  \multicolumn{1}{l}{} \\
$\hat{\sigma}_2$ &
  $1$ &
  \multicolumn{1}{c}{$-1$} &
  \begin{tabular}[c]{@{}c@{}}$\begin{pmatrix}     \frac{1}{\sqrt{2}}  \\     \frac{i}{\sqrt{2}}     \end{pmatrix}  $\end{tabular} &
  \multicolumn{1}{c}{\begin{tabular}[c]{@{}c@{}}$\begin{pmatrix}     \frac{1}{\sqrt{2}}  \\     \frac{-i}{\sqrt{2}}     \end{pmatrix}  $\end{tabular}} &
   $\ket{R}$&
  \multicolumn{1}{c}{$\ket{L}$} & $\bra{R}\hat{\rho}\ket{R} - \bra{L}\hat{\rho}\ket{L}$
   \\
\multicolumn{1}{l}{} &
  \multicolumn{1}{l}{} &
  \multicolumn{1}{l}{} &
  \multicolumn{1}{l}{} &
  \multicolumn{1}{l}{} \\
$\hat{\sigma}_3$ &
  $1$ &
  \multicolumn{1}{c}{$-1$} &
  \begin{tabular}[c]{@{}c@{}}$\begin{pmatrix}     1  \\     0     \end{pmatrix}  $\end{tabular} &
  \multicolumn{1}{c}{\begin{tabular}[c]{@{}c@{}}$\begin{pmatrix}     0  \\     1     \end{pmatrix}  $\end{tabular}} &
  $\ket{H}$ &
  \multicolumn{1}{c}{$\ket{V}$} & $\bra{H}\hat{\rho}\ket{H} - \bra{V}\hat{\rho}\ket{V}$
\end{tabular}
\caption{Mathematical representation of an $N=2$ state tomography setup.}
\label{tab:stokes_polarimetry_setup}
\end{table}

\endgroup

\begin{figure}[htbp]
    \centering
    \includegraphics[width=0.99\linewidth]{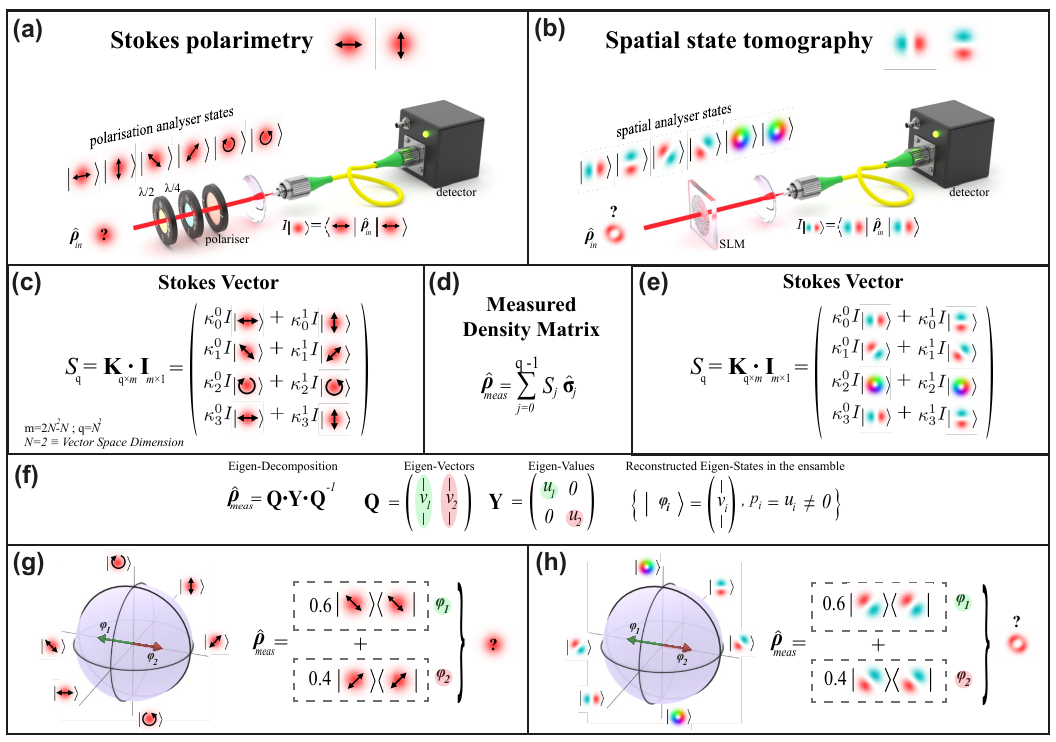}
    \caption{Stokes polarimetry analogy for $N=2$. (a, b) Experimental setups for polarisation and spatial tomography, using waveplates/polariser or an SLM respectively. (c, e) Intensity measurements are converted into a Stokes vector $\mathbf{S}$. (d, f--h) The density matrix is reconstructed and eigendecomposed to recover the pure states in the mixture. Adapted from \cite{si_ploschner_spatial_2022}.}
    \label{fig:stokes_analogy}
\end{figure}

While the previous table sets out the theory for a generic $N=2$ tomographic measurement, Figure~\ref{fig:stokes_analogy} illustrates its experimental realisation, comparing Stokes polarimetry with the spatial $N=2$ analogue side by side. The tomographic framework is unchanged: only the degree of freedom under measurement differs (polarisation versus space). In both, (a) and (b), the same unknown input mixed state, denoted $\hat{\rho}_{{in}}$, is prepared in the basis appropriate to each degree of freedom. The six analyser states are projected sequentially using a $\lambda/2$ plate, a $\lambda/4$ plate, and a polariser in conventional Stokes polarimetry, and using an SLM in SST. For each analyser state, the transmitted intensity is recorded with a single-pixel detector formed by a lens, single-mode fibre (SMF), and photodiode, equivalently to Eq.~\eqref{eq:intensity_probability}. Once all projections are complete, the mixed state is reconstructed following the formalism above: the measured intensities are converted into the Stokes vector $\mathbf{S}$ (c and e), $\hat{\rho}_{in}$ is recovered from $\mathbf{S}$ (d), and its eigenstates are extracted by eigendecomposition (f). Finally, (g) and (h) show that the unknown mixture is an incoherent superposition of two pure states with intensity weights 3:2---the same mixture encoded in polarisation and in space.

\supplnote{sn:mmf-mtm}{SST concept for MMF characterisation}

\begin{figure}[htbp]
    \centering
    \includegraphics[width=0.97\linewidth]{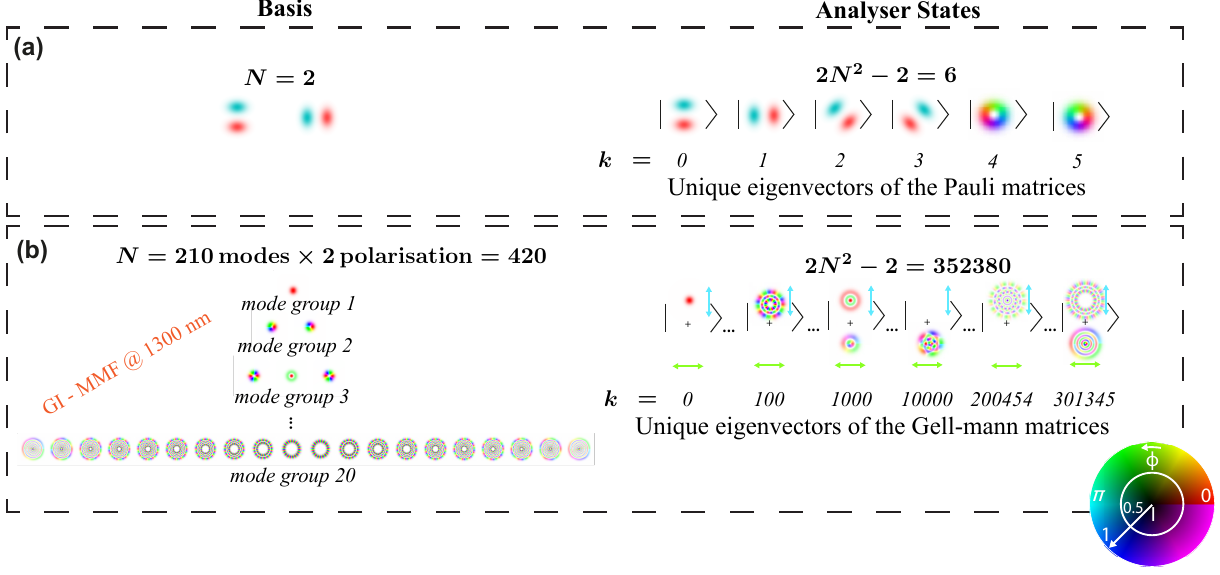}
    \caption{Input analyser states for MMF characterisation. (a) Six analyser states for Stokes polarimetry ($N=2$). (b) High-dimensional SST for the OM1 GI-MMF ($N=420$), requiring $2N^2-N=352{,}380$ pairwise mode-interference projections at four phase offsets, plus single-mode terms.}
    \label{fig:high_stokes}
\end{figure}

The SST concept applied to MMF characterisation remains the same but extends the dimensionality to the number of spatial modes supported by the fibre-under-test. For the particular case presented in the main article, the employed fibre is a OM1 GI-MMF supporting 20 mode groups at 1300~nm and two polarisation states. This is equivalent to 420 modes across both polarisations (210 each) since they are coupled by the fibre. Analoguously to the previous SST example for $N=2$, this scenario is expanded as shown in Figure~\ref{fig:high_stokes}.

A distinctive feature of MMF characterisation employing SST is that the fibre itself does not constitude a mixed state, it produces a wavelength dependent spatial transformation to the input light. For a coherent source at a single wavelength, the output is a pure state—a linear combination of the supported basis modes weighted by fibre-induced mode mixing, which varies with input mode and modal dispersion. It can be infered that if we can probe the fibre with a set of orthogonal basis at the input and untagle the mixing using above SST formalism we can construct a TM row by row.

\begin{figure}[htbp]
    \centering
    \includegraphics[width=0.97\linewidth]{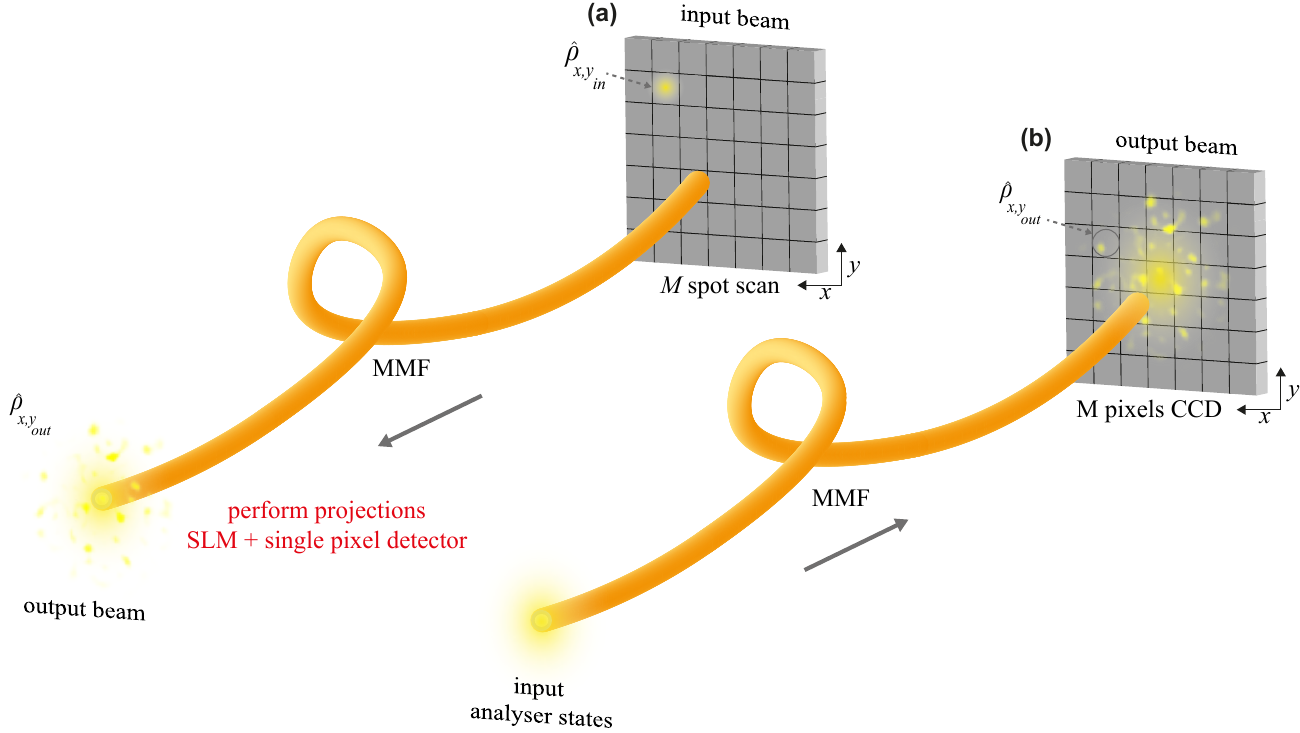}
    \caption{SST as an MMF characterisation technique. (a) Each row of the TM is obtained by performing high-dimensional tomography at the fibre output for $M^2$ orthogonal input illuminations. (b) Injecting the $2N^2-N$ analyser states at the input and recording $I(x,y,k)$ with a camera parallelises $M^2$ virtual tomographies, yielding one density matrix per pixel.}
    \label{fig:combined_sst}
\end{figure}

A natural probing basis is the spot basis: $M^2$ illuminations on the proximal facet, arranged for simplicity as an $M \times M$ array (Figure~\ref{fig:combined_sst}(a)). For each input spot $(x,y)$, the fibre produces an unknown output field at the distal facet. Applying the same SST protocol as in the Stokes analogy—analyser-state projection with an SLM and readout by a single-pixel detector (lens, SMF, photodiode)—recovers the modal superposition at the output, i.e.\ the complex coefficients that map that input spot to the distal field (or, for a partially coherent source, the incoherent mixture of modal superpositions). Repeating for all $M^2$ spots yields a TM mapping $N$ input modes to $M^2$ output spots, albeit phase-unresolved between spots. The cost is prohibitive: each row requires a full $2N^2-N$ projection sequence, so the conventional sequential approach scales as $M^2 \times (2N^2-N)$ acquisitions.

We invert the problem by exploiting optical reciprocity and parallel detection with a single camera (Figure~\ref{fig:combined_sst}(b)). Instead of analysing the output field separately for each input spot, the SLM injects the $2N^2-N$ analyser states at the proximal facet; one camera frame per analyser state records the intensity at every pixel $I(x,y,k)$. The pixel grid provides an orthogonal spatial sampling basis: each pixel measures an intensity expectation in a distinct region of the output plane, so $M^2$ virtual tomographies run in parallel. Processing the $k$-stack at each pixel $(x,y)$ yields a local density matrix $\hat{\rho}_{x,y}$ and its primary eigenstate—the input modal superposition that focuses power at that pixel, as described in the main article. The same $N \leftrightarrow M^2$ mapping as in Figure~\ref{fig:combined_sst}(a) is obtained in a single tomographic pass rather than $M^2$ sequential measurements.

\subsubsection*{MTM calculation}

\begin{figure}[htbp]
    \centering
    \includegraphics[width=0.97\linewidth]{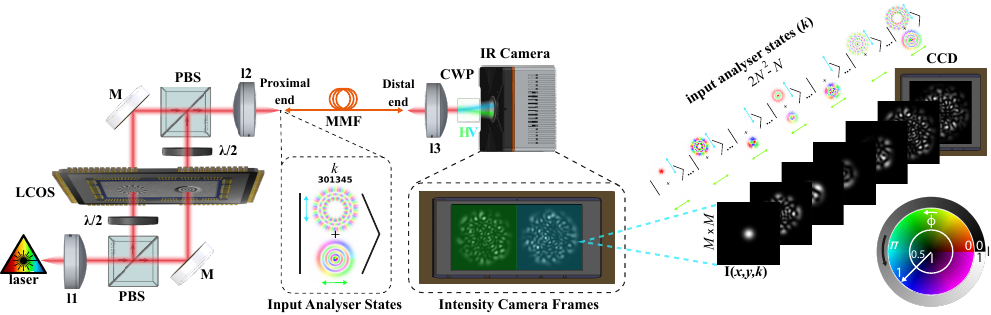}
    \caption{Polarisation-resolved data layout for MTM retrieval. Each acquired frame is divided into horizontal and vertical regions by the Wollaston prism. The two region of interest (ROIs) are processed independently to build $\mathrm{TM}_{\mathrm{H}}$ and $\mathrm{TM}_{\mathrm{V}}$ in the output spot basis.}
    \label{fig:launching_analyser_states}
\end{figure}

Figure~\ref{fig:launching_analyser_states} highlights a key feature of the retrieval pipeline: each camera frame is split into horizontal and vertical components by the distal Wollaston prism, and the two halves are stored and processed independently. This polarisation diversity yields two spot-domain transmission matrices, $\mathrm{TM}_{\mathrm{H}}$ and $\mathrm{TM}_{\mathrm{V}}$, each mapping the $N$ input modes---$N/2$ spatial modes per polarisation---to the $M^2$ camera pixels.

\clearpage

\begin{figure}[htbp]
    \centering
    \includegraphics[width=0.85\linewidth]{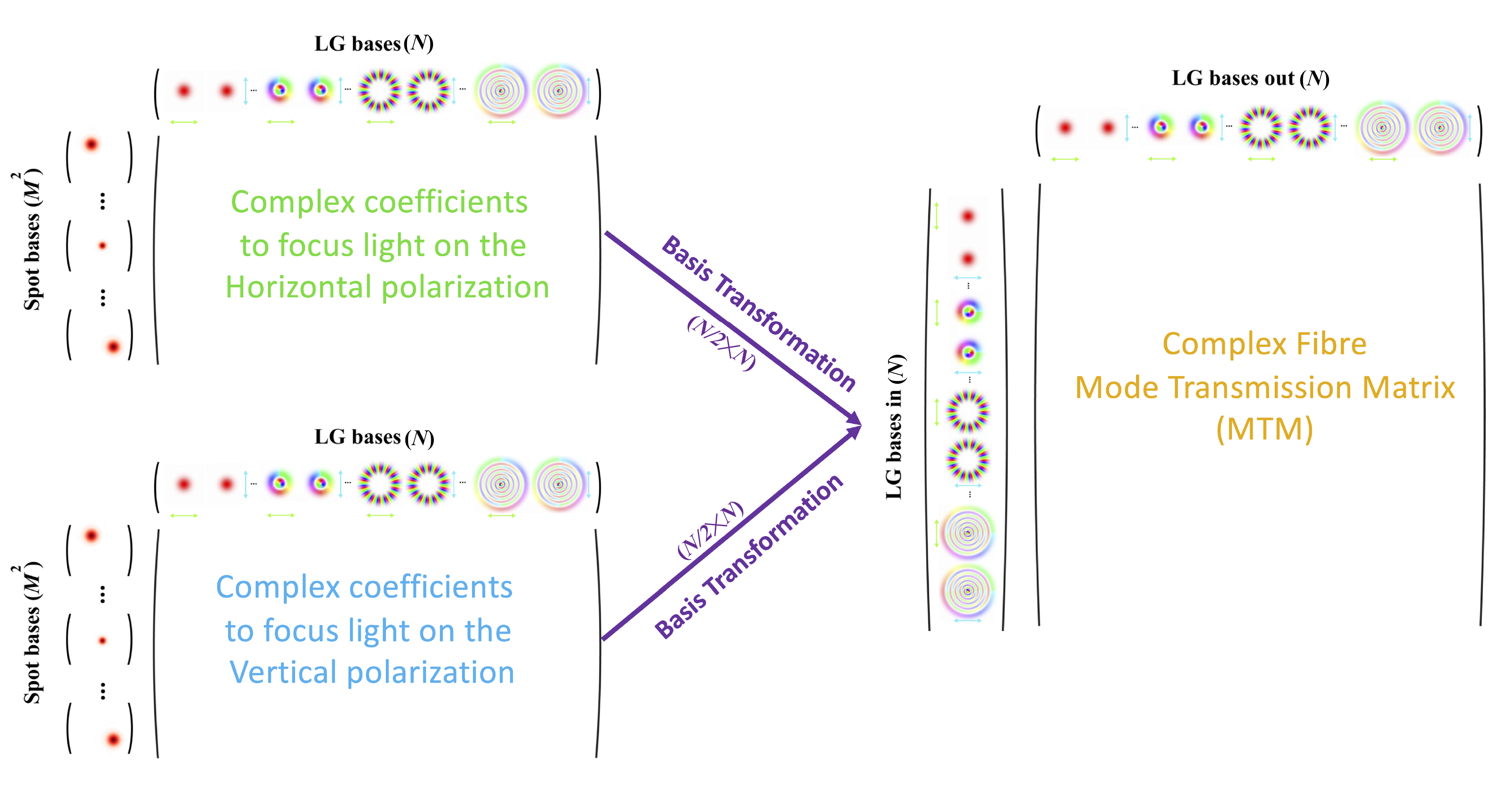}
    \caption{Complete polarisation-resolved MTM assembly. For each polarisation (H, V), the phase-corrected spot-domain $\mathrm{TM}$ is transformed to the LG input basis. The two blocks are interleaved into a single $N \times N$ MTM that fully describes spatial and polarisation coupling through the MMF.}
    \label{fig:complete_MTM_SST}
\end{figure}

Each spot-domain matrix is initially known only up to an unknown phase per pixel. The phase-locking algorithm described in the main manuscript (Methods~\ref{meth:phase}) recovers the relative phase between neighbouring pixels within the same speckle grain, propagating a globally consistent phase across the full $M^2$ spot basis without an external reference. Once phase-corrected, each $\mathrm{TM}$ is transformed from the output pixel basis to the Laguerre--Gaussian (LG) input basis of dimension $N/2$, giving $\mathrm{MTM}_{\mathrm{H}}$ and $\mathrm{MTM}_{\mathrm{V}}$. Interleaving the two polarisation blocks assembles the complete $N \times N$ polarisation-resolved MTM that describes coupled spatial and polarisation propagation through the fibre (Figure~\ref{fig:complete_MTM_SST}). This eliminates the redundant oversampling of the spot representation and expresses the measured response entirely in the modal basis used for wavefront shaping. As detailed in~\cite{si_ploschner_seeing_2015, si_mahalati_resolution_2013}, the condition $M^2 \geq 4N$ is sufficient to retrieve the complete MTM; downsampling of the camera grid can be applied when this oversampling criterion is met.

\supplnote{sn:lab-setup}{Laboratory diagram, alignment and data handling}

\begin{figure}[htbp]
    \centering
    \includegraphics[width=0.99\linewidth]{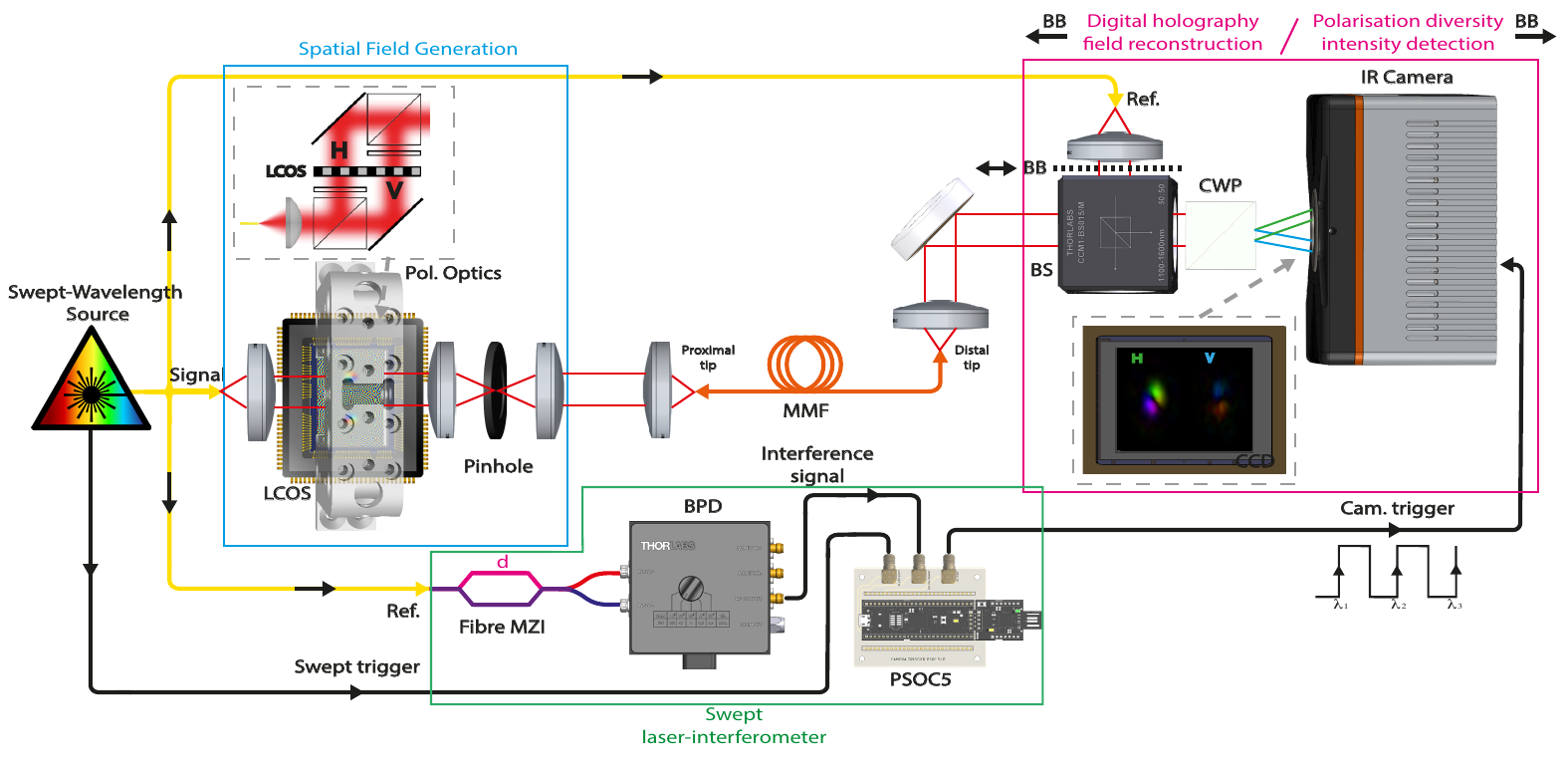}
    \caption{Extended laboratory setup corresponding to main text Fig.~1(a). Blue: proximal wavefront-shaping stage (SLM, polarisation optics, and lenses). Pink: distal detection with Wollaston prism for polarisation diversity. Green: swept-laser trigger path. Left beam blocker (BB): off-axis digital holography reference arm; right BB: SST intensity-only detection.}
    \label{fig:setup}
\end{figure}

Figure~\ref{fig:setup} expands Fig.~1(a) of the main text with colour-coded stages. The apparatus is designed for two characterisation stages on the same optical bench: referenceless SST, in which the holographic reference arm is blocked and only distal intensities are recorded; and off-axis digital holography (DH), in which the reference arm is unblocked to recover complex output fields for benchmarking. Evaluation metrics, the swept-laser protocol, and phase locking are defined in the main article Methods~\ref{meth:OI}, \ref{meth:SCS}, and \ref{meth:phase}. This note records the alignment practice, acquisition layout, and high-$N$ storage and processing workflow required to reproduce the measurements.

\subsubsection*{Optical stages}

The proximal stage (blue) generates arbitrary spatial and polarisation fields at the fibre input. A swept-frequency laser (New Focus Venturi TLB-8800, 80~nm tuning range centred at 1300~nm) feeds an LCoS SLM (Meadowlark P1920) placed near the Fourier plane of the proximal optics. Waveplates, polarising beam splitters, and a Gerchberg--Saxton algorithm provide independent horizontal (H) and vertical (V) control of the launched field. The fibre-under-test (displayed in orange) is a 2~m OM1 graded-index MMF (62.5~\textmu m core) supporting 20 near-degenerate mode groups at 1300~nm, equivalent to $N=420$ spatial and polarisation modes. The distal stage (pink) images the fibre facet onto a Xenics Cheetah InGaAs camera (640$\times$512 pixels, 20~\textmu m pitch). A calcite Wollaston prism (CWP) separates H and V onto opposite halves of the sensor so that each polarisation is processed independently. The green path implements swept-laser interferometry: a balanced photodiode converts the interferogram frindges into TTL triggers that synchronise wavelength-resolved camera frames during SCS emulation (main Methods, swept-laser protocol)~\cite{si_MaestreMorote2026pySSTri}.

\subsubsection*{Proximal alignment}

Accurate proximal coupling is critical because the analyser states are constructed in the LG modal basis; misalignment or uncorrected aberrations degrade mask fidelity and propagate into every projection. Alignment proceeds in two steps with the MMF temporarily replaced by a single-mode fibre (SMF) and a power meter at the proximal facet.

First, a blazed grating displayed on the SLM deflects the zero-order away from the fibre and couples only the first diffraction order. The SMF position is adjusted manually until maximum throughput is achieved, placing the fibre as close as practicable to the SLM Fourier plane while minimising residual beam tilt that cannot be corrected digitally. Higher diffraction orders are rejected by the fibre cladding (single lens system) or, where used, a pinhole (4$f$ configuration system).

Second, residual aberrations from the SLM optics and any manual misalignment are corrected by optimising Zernike polynomials (degree $\leq 6$) displayed on the SLM while monitoring coupled power into the SMF. Optimisation runs until convergence; the MMF is then reinstalled without moving the SLM carriage or proximal optics, preserving the calibrated wavefront.

\subsubsection*{Distal end and dual detection paths}

For SST measurements the reference arm is blocked and only intensity is recorded at the camera. No distal phase correction is required because the tomographic reconstruction and phase-locking algorithm (main Methods~\ref{meth:phase}) recover the complex TM from intensity data alone. For DH benchmarking the reference arm is unblocked, forming off-axis interferograms on both polarisation halves of the sensor. Residual aberrations in the reference and imaging path are removed in software following the calibration procedures in~\cite{si_carpenter_digholo_2022, si_mounaix_control_2019}. Switching between SST and DH therefore requires only repositioning the beam block; the fibre, camera, and polarisation optics remain fixed.

\subsubsection*{Tomographic data layout}

Raw SST data are organised as a three-dimensional intensity array $I(x,y,k)$, where $k$ indexes the $m=2N^2-N$ input analyser states and $(x,y)$ spans the camera ROI. For the full 20 mode-group dual polarisation measurement ($N=420$), each $k$-frame yields two $200\times200$ pixel, 16-bit ROIs (one per polarisation), giving $\sim$110~GB on disk for a single LCS acquisition. Frames are stored so that each pixel can be loaded independently along $k$, keeping RAM usage bounded during post-processing.

For SCS emulation the swept-laser trigger records $D=118$ wavelength-resolved frames per analyser state ($\lambda = 1280$--1320~nm), equivalent to $D$ independent monochromatic measurements acquired in one experimental run. Because the camera is much faster than the LCoS refresh rate, the total SST acquisition time is mostly set by hologram generation, not by the number of triggered wavelengths. The spectrally resolved archives scale to $\sim$10~TB and are kept on network-attached storage (NAS). Emulating a desired source bandwidth $\Delta\lambda_S$ in post-processing requires reading and summing intensity frames along the wavelength axis for each $k$ (main Methods~\ref{meth:SCS}); in our configuration, NAS read bandwidth dominates this step ($\sim$19~h), whereas the on-line tomographic acquisition duration is unchanged.

\subsubsection*{Sparse reconstruction}

At $N=420$, the dense Gell--Mann operators $\hat{\sigma}_j$ and weighting matrix $\mathbf{K}$ exceed available RAM if assembled naively. Both structures are highly sparse: most $\hat{\sigma}_j$ entries are zero, and $\mathbf{K}$ contains only the eigenvalues $\kappa_j^r$ associated with each retained analyser state. Exploiting this sparsity is essential for practical reconstruction of Eqs.~\eqref{eq:rho_recon_SI} and~\eqref{eq:stokes_matrix_SI}. Sparse implementations of the Stokes-vector and density-matrix recovery are provided in the GitHub repository \url{https://github.com/MarKo7s/SST_MMF_characterization}; post-processing of the full 110~GB LCS dataset is GPU-accelerated and completes in $\sim$30~min on the workstation described in the main text.

\supplnote{sn:lcs-extra}{Long coherence source --- additional figures}

\subsubsection*{Short acquisition-time SST}

\begin{figure}[htbp]
    \centering
    \includegraphics[width=0.97\linewidth]{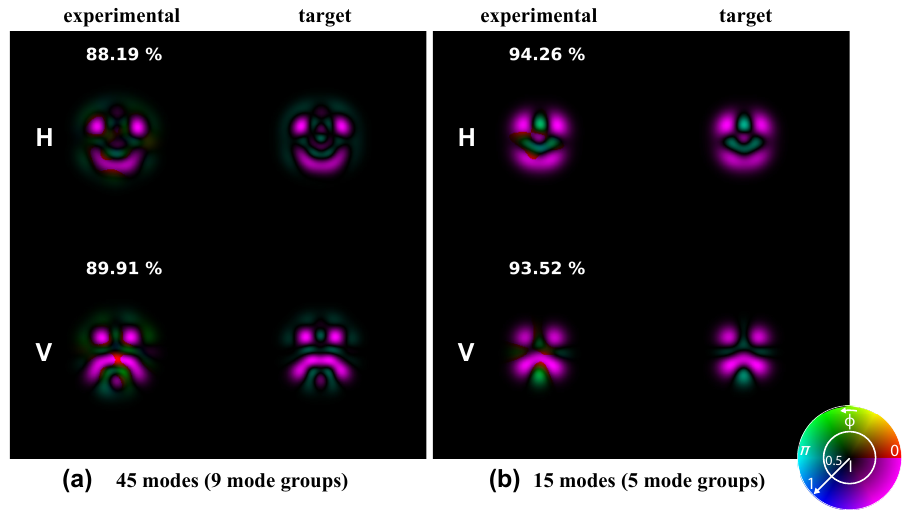}
    \caption{Effect of acquisition time on LCS beam shaping. Partial-mode-group SST (5 and 9 mode groups; 17~min and 2~h acquisitions) yields flat reconstructed phases, whereas the full 420-mode run acquired over 54~h exhibits phase ripples attributed to mechanical and thermal drift rather than SST bias.}
    \label{fig:short_sst}
\end{figure}

Main text Fig.~2 shows phase ripples on some LCS targets despite favourable singular values (main Methods~\ref{meth:SVD}). We attribute this to 54~h drift, not SST bias. The results shown in Fig.~\ref{fig:short_sst} correspond to SST MTM measurements on 5 and 9 mode groups only (17~min for 30 modes; 2~h for 90 modes) which removes the phase artefacts seen on the long tomographic measurement.

\subsubsection*{Intensity overlap figure}

\begin{figure}[htbp]
    \centering
    \includegraphics[width=0.97\linewidth]{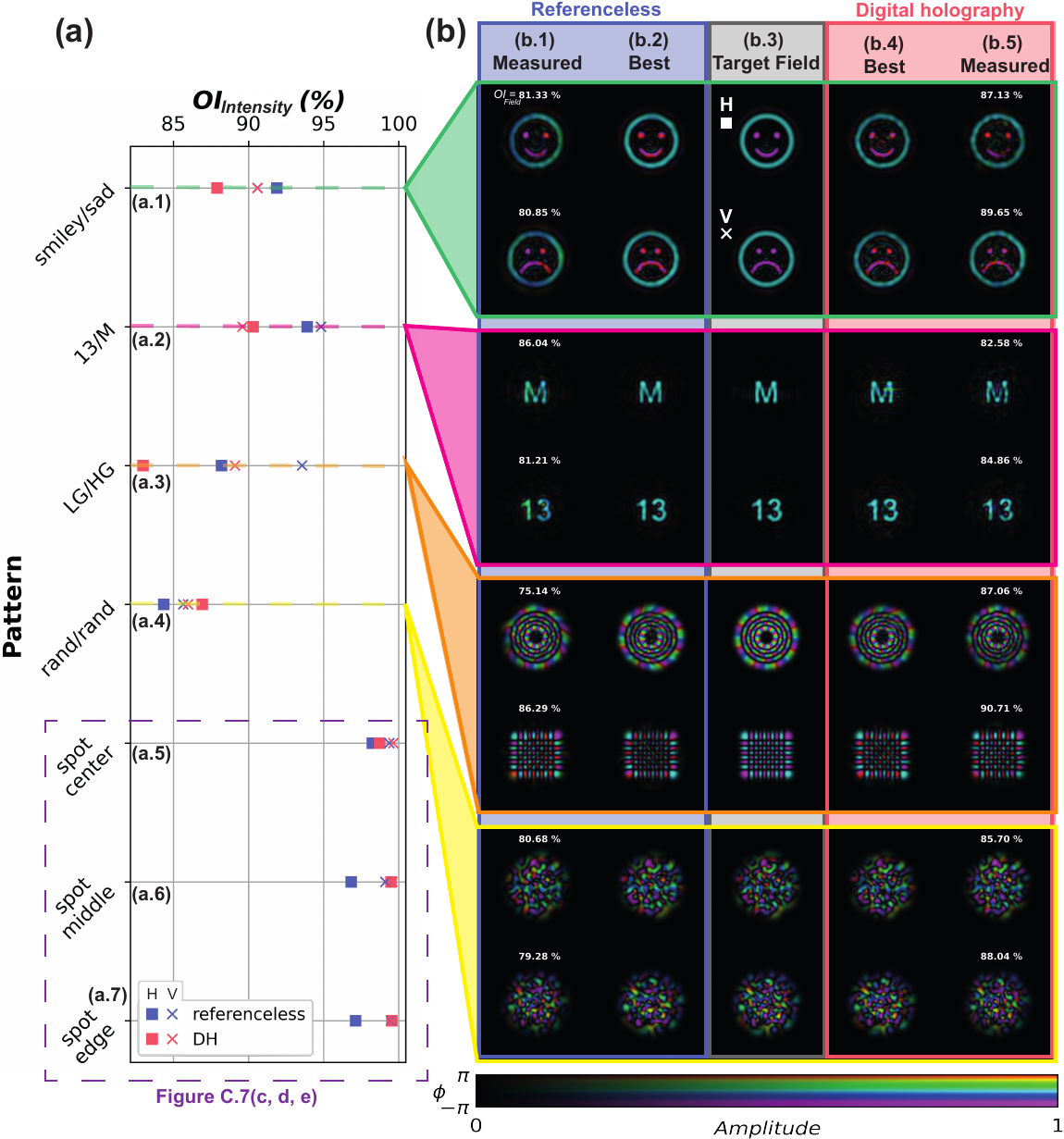}
    \caption{Comparison of intensity and complex overlap metrics for LCS beam shaping. (a) Intensity overlap $OI_{\mathrm{I}}$ for four target patterns, showing closer agreement between MTM$_{SST}$ and visual field quality. (b) Target fields, theoretical best-case patterns, and DH-measured outputs; complex $OI$ values are annotated above each measured field.}
    \label{fig:LCS_intensity}
\end{figure}

Figure~\ref{fig:LCS_intensity} plots the intensity overlap $OI_{\mathrm{I}}$ alongside the complex overlap $OI$ (both defined in the main Methods~\ref{meth:OI}). $OI_{\mathrm{I}}$ tracks visual quality more closely because it neglects phase artefacts that can arise from the holographic reference arm or from mechanical and thermal drift during a long tomographic acquisition. In this comparison, MTM$_{SST}$ outperforms MTM$_{DH}$ for those patterns whose measured phase is clearly non-flat: the phase errors reduce the complex $OI$ even when intensity delivery remains strong.

\supplnote{sn:fibre-lambda}{Fibre temporal response and correlation bandwidth}

The temporal and spectral properties of the tested OM1 fibre can be examined by performing the Fast Fourier Transform (FFT) across the frequency/wavenumber axis of the spectrally resolved MTM, which was measured using an external reference and off-axis digital holography. By grouping the modes into near-degenerate mode groups and computing the FFT, the impulse response of each mode group is obtained, revealing the time delay associated with each group. 

\begin{figure}[htbp]
    \centering
    \includegraphics[width=0.97\linewidth]{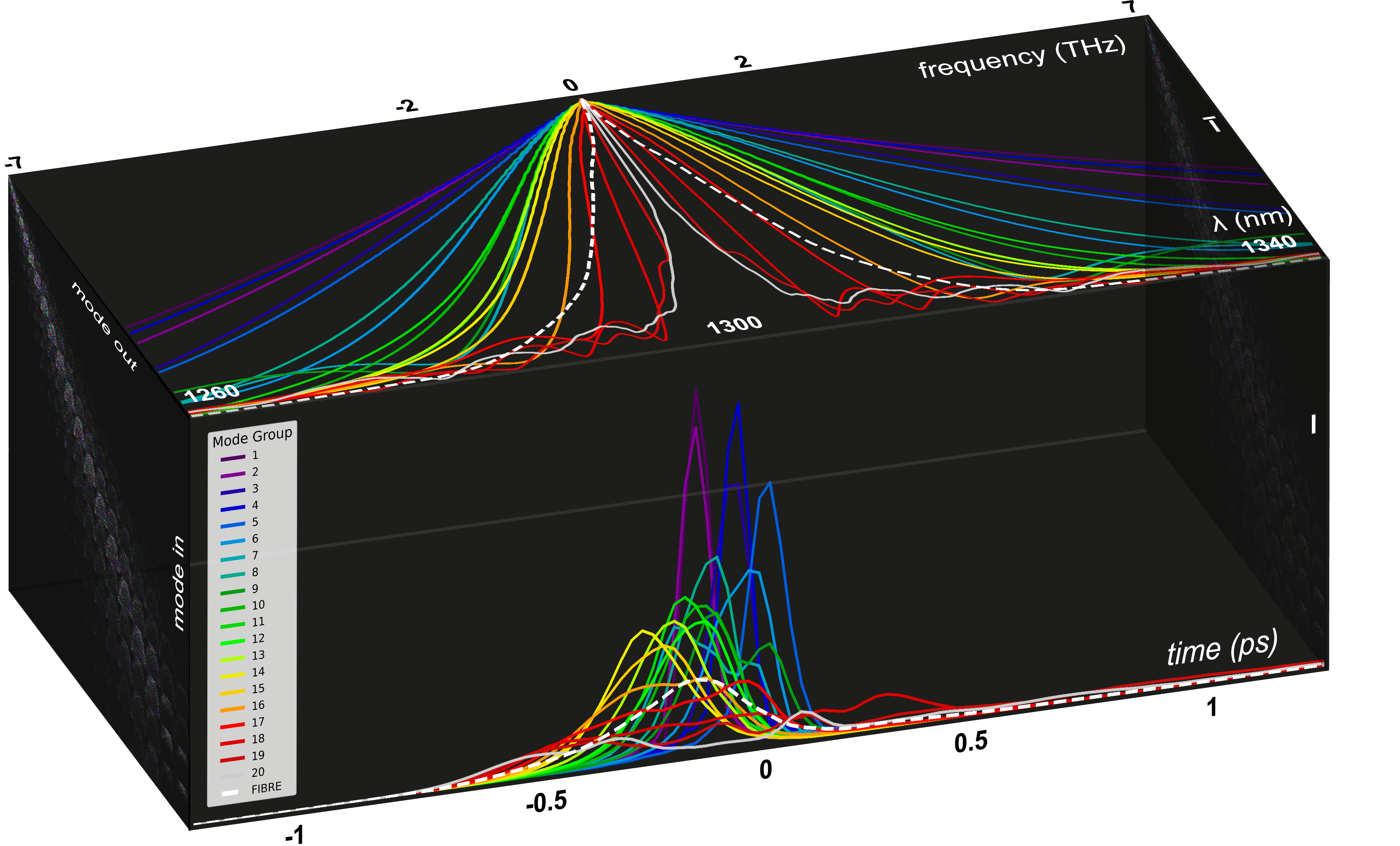}
    \caption{Spectral and temporal characterisation of the 2~m OM1 GI-MMF from a spectrally resolved MTM measured by off-axis digital holography. (a) Normalised impulse response of each near-degenerate mode group, showing minimal inter-group delay. (b) Generalised frequency correlation of the full transmission matrix, yielding a fibre correlation bandwidth $\Lambda \approx 10$~nm used to interpret the SCS results in the main text and Note~\ref{sn:scs-extra}.}
    \label{fig:OM1_spectral_temporal_results}
\end{figure}

Figure~\ref{fig:OM1_spectral_temporal_results} illustrates the spectral and temporal results for the OM1 fibre. The 2~m custom patch cord exhibits a very narrow total impulse response, meaning each individual mode group travels very close to each other with minimal inter-group modal dispersion. This temporal behaviour translates into a broad correlation bandwidth in the frequency domain, with a total fibre correlation bandwidth $\Lambda \approx 10$~nm. 

\supplnote{sn:scs-extra}{Short coherence source --- additional figures}

\subsubsection*{Eigenvalue maps}

\begin{figure}[htbp]
    \centering
    \includegraphics[width=0.65\linewidth]{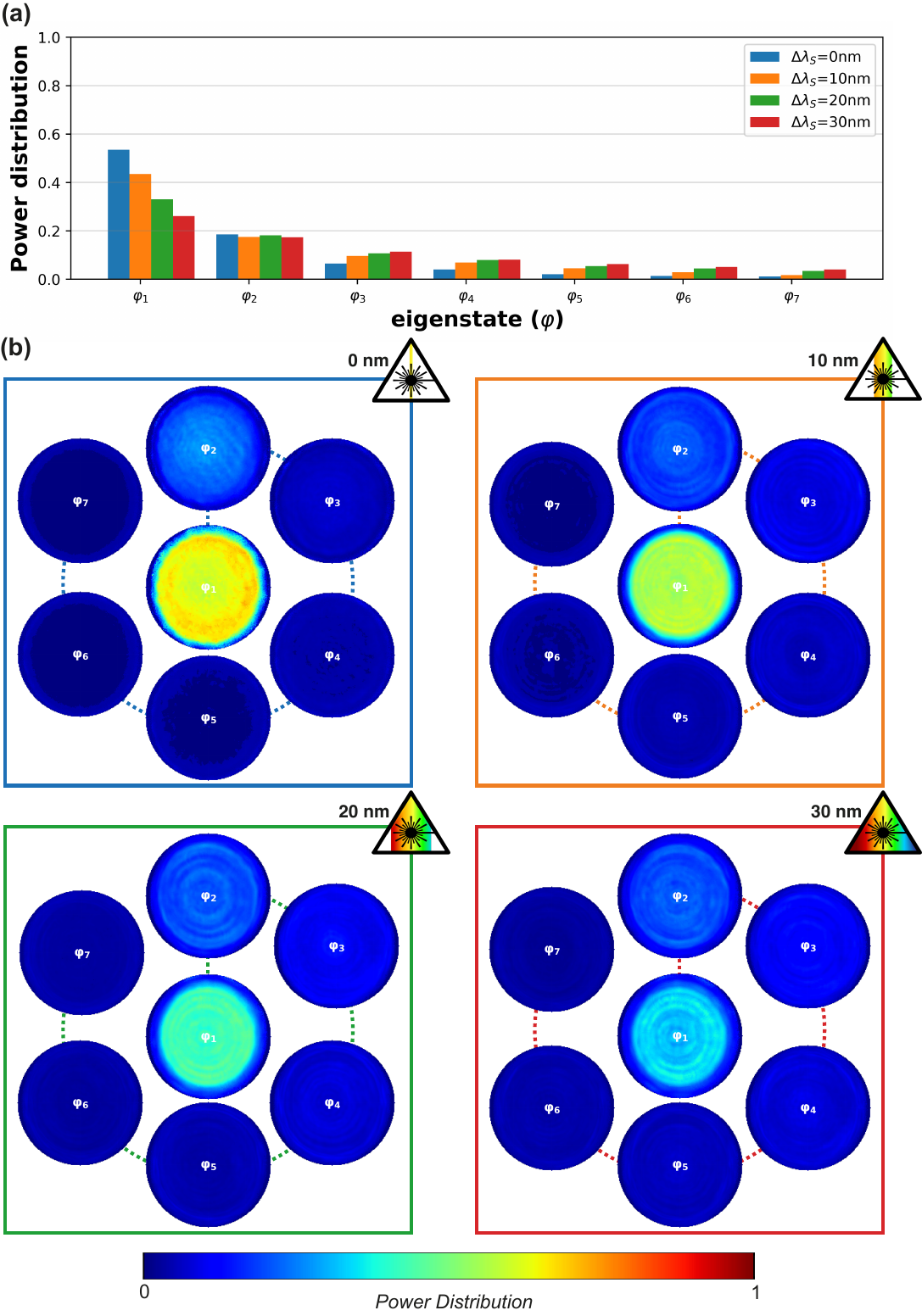}
    \caption{Spatial distribution of the first seven density-matrix eigenvalues for emulated source bandwidths $\Delta\lambda_S = 0$, 10, 20, and 30~nm. Left column: mean eigenvalue maps; right columns: per-pixel distributions. As $\Delta\lambda_S$ increases, $\varphi_1$ concentrates in the core while secondary eigenstates gain weight towards the cladding, reflecting spatial decoherence across mode groups.}
    \label{fig:eigenstates}
\end{figure}

Figure~\ref{fig:eigenstates} displays the spatial distribution of the first seven eigenstates at experimentally emulated source bandwidths of $\Delta\lambda_S = 0$, 10, 20, and 30~nm. The thresholding and positive semi-definite (PSD) projection procedures follow the methodology outlined in Note~\ref{sn:sst-theory}. The absolute scales are indicative, but the spatial trends clearly reflect mode-dependent loss and decorrelation. As the bandwidth increases, the primary eigenstate ($\varphi_1$) tends to concentrate power in the core (lower-order modes), while the secondary eigenstates begin to capture the decorrelated power spreading towards the cladding (higher-order modes).

\subsubsection*{Secondary MTMs}

\begin{figure}[htbp]
    \centering
    \begin{minipage}[t]{0.48\linewidth}
        \centering
        \includegraphics[width=\linewidth]{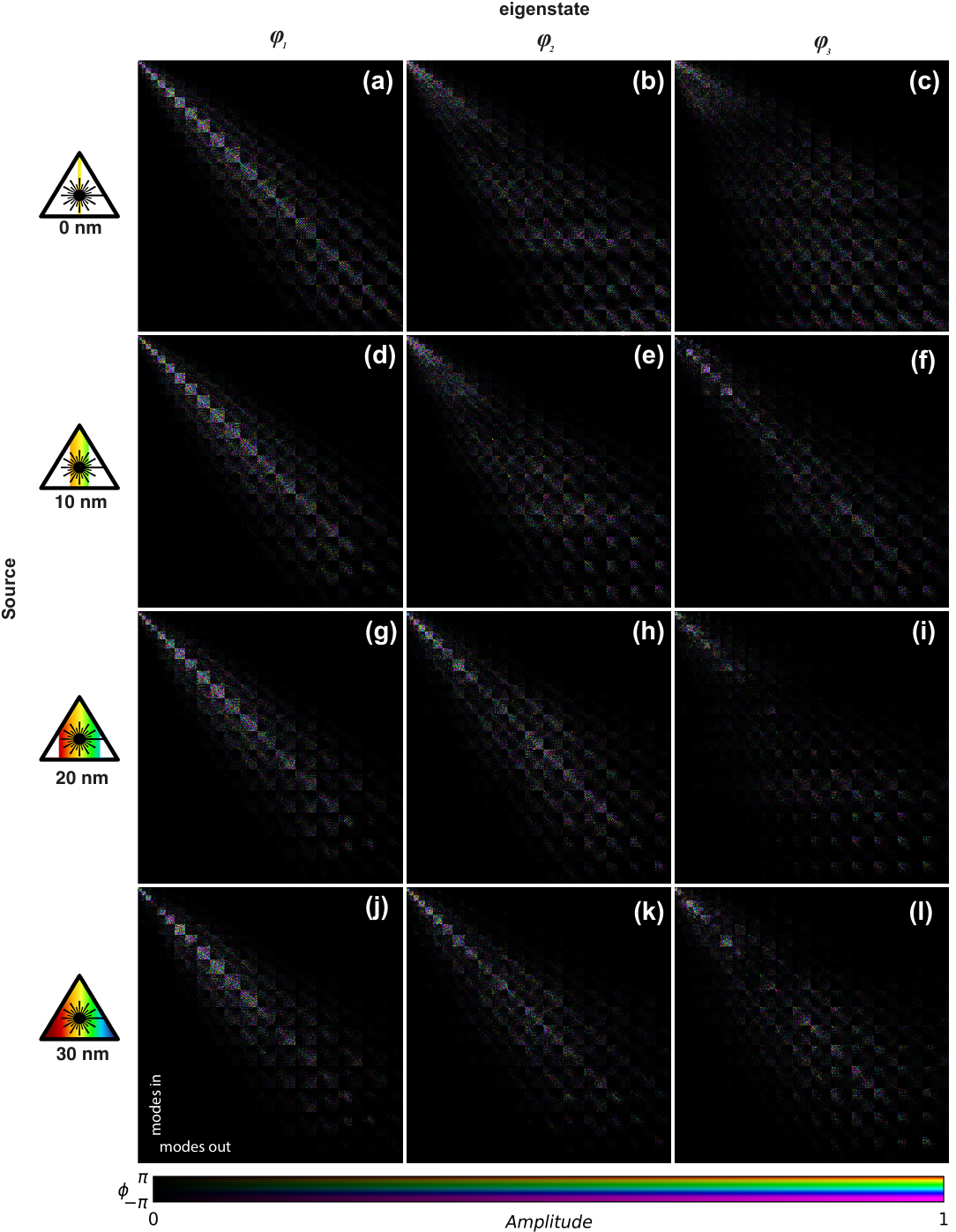}\\[0.4em]
        {\footnotesize (i)}
    \end{minipage}
    \hfill
    \begin{minipage}[t]{0.48\linewidth}
        \centering
        \includegraphics[width=\linewidth]{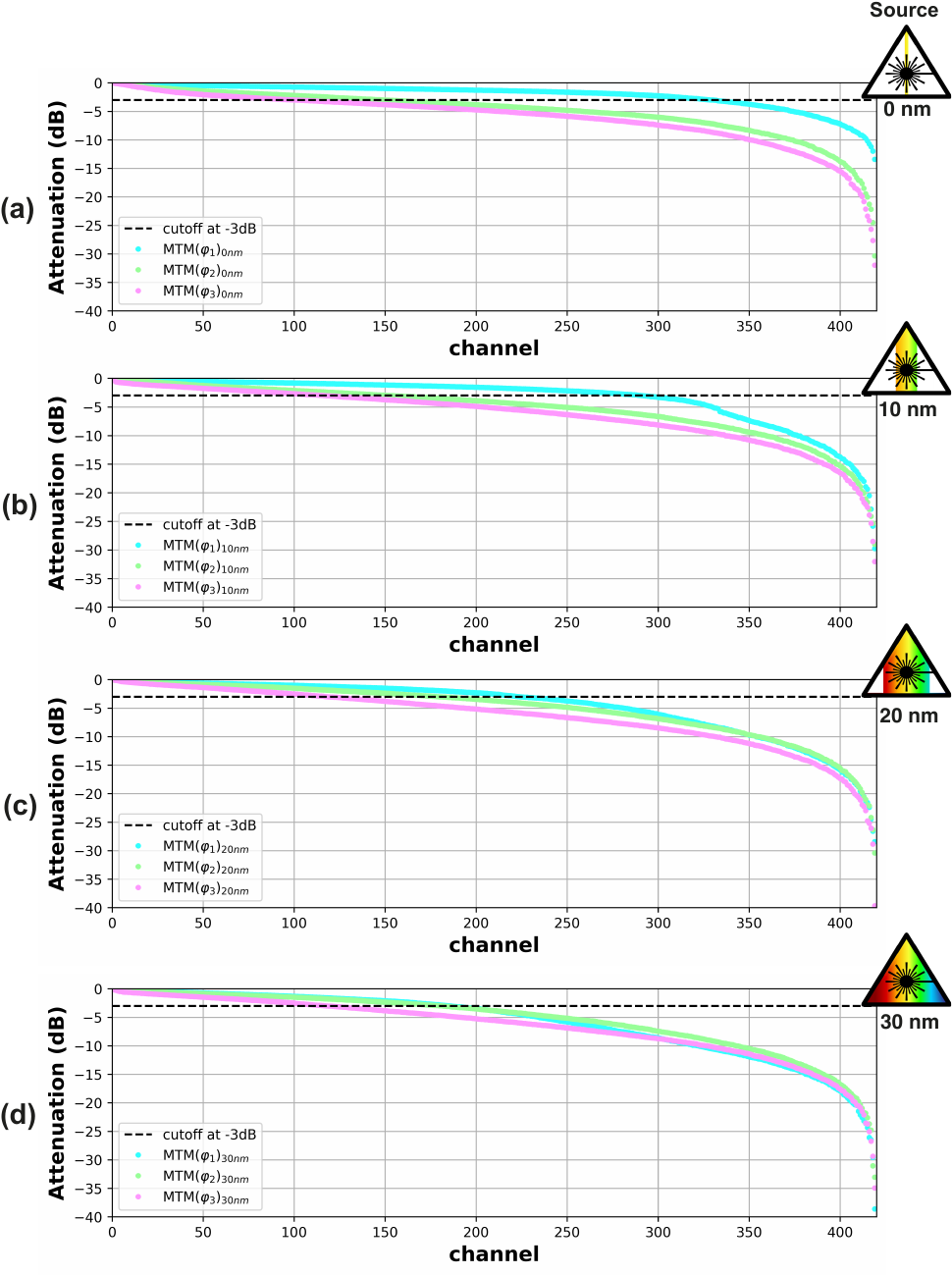}\\[0.4em]
        {\footnotesize (ii)}
    \end{minipage}

    \caption{%
    (i)~Mode transmission matrices MTM($\varphi_i$) for eigenstates $i=1,2,3$ and emulated bandwidths $\Delta\lambda_S = 0$, 10, 20, and 30~nm. At $\Delta\lambda_S=0$, only $\varphi_1$ is physical; $i\geq2$ matrices are noise. At $\Delta\lambda_S=30$~nm, higher-order mode groups absent from MTM($\varphi_1$) reappear in MTM($\varphi_2$) and MTM($\varphi_3$).
    (ii)~Normalised singular values of MTM($\varphi_i$)$_{\Delta\lambda_S}$ for $i=1,2,3$ and $\Delta\lambda_S = 0$, 10, 20, and 30~nm. Channel attenuation in $\varphi_1$ increases with bandwidth as fewer modes remain mutually coherent; secondary eigenstates retain usable channels for decorrelated modal subsets.}
    \label{fig:MTMs_SVDs}
\end{figure}

The MTM per eigenstate (MTM$\varphi_i$) are reconstructed by collapsing all pixels associated to a eigenstate into the MTM as outline in Note~\ref{sn:mmf-mtm}. Figures~\ref{fig:MTMs_SVDs} illustrate the secondary eigenstates and their corresponding singular values for $i=1,2,3$. In the monochromatic case ($\Delta\lambda_S=0$), the system is fully spatially coherent, meaning the density matrix possesses only a single non-zero eigenstate ($\varphi_1$). Consequently, any channels appearing in $i\geq2$ are merely experimental artefacts or noise, as discussed in the main Methods~\ref{meth:SVD}. 

However, as the source bandwidth increases, spatial decoherence emerges. For instance, at $\Delta\lambda_S=30$~nm, the highest eigenstate ($\varphi_1$) can no longer maintain a phase relationship across all modes, causing certain higher-order mode groups to drop out of MTM($\varphi_1$). These "missing" modes are instead captured by the secondary eigenstates. MTM($\varphi_2$) and MTM($\varphi_3$) thus carry these sideband and cladding-weighted solutions, representing mutually incoherent components of the mixed state that can still be utilised for specific wavefront shaping tasks.

% Both figures side to side

% Independent Figures

% \begin{figure}[htbp]
%     \centering
%     \includegraphics[width=0.95\linewidth]{figures/second_MTM.pdf}
%     \caption{Mode transmission matrices MTM($\varphi_i$) for eigenstates $i=1,2,3$ and emulated bandwidths $\Delta\lambda_S = 0$, 10, 20, and 30~nm. At $\Delta\lambda_S=0$, only $\varphi_1$ is physical; $i\geq2$ matrices are noise. At $\Delta\lambda_S=30$~nm, higher-order mode groups absent from MTM($\varphi_1$) reappear in MTM($\varphi_2$) and MTM($\varphi_3$).}
%     \label{fig:MTMs}
% \end{figure}

% \begin{figure}[htbp]
%     \centering
%     \includegraphics[width=0.95\linewidth]{figures/second_svd.pdf}
%     \caption{Normalised singular values of MTM($\varphi_i$)$_{\Delta\lambda_S}$ for $i=1,2,3$ and $\Delta\lambda_S = 0$, 10, 20, and 30~nm. Channel attenuation in $\varphi_1$ increases with bandwidth as fewer modes remain mutually coherent; secondary eigenstates retain usable channels for decorrelated modal subsets.}
%     \label{fig:SVDs}
% \end{figure}

\subsubsection*{Beam shaping: Random Superposition of modes MTM($\varphi_1$) - Additional results}

\begin{figure}[htbp]
    \centering
    \includegraphics[width=0.95\linewidth]{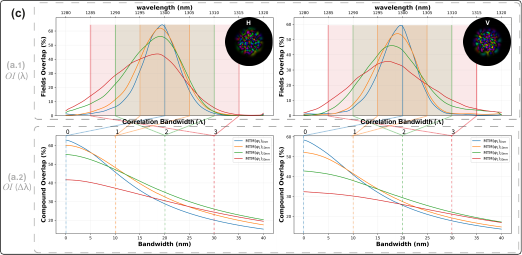}
    \caption{Broadband beam shaping with a random amplitude--phase superposition of all fibre modes. Overlap $OI(\lambda)$ and integrated $OI(\Delta\lambda)$ for each retrieved MTM$_{\Delta\lambda_S}$, reproducing the $\sqrt{2}$ bandwidth-enhancement trend observed for the smiley/sad target in the main text. Optimal performance occurs when the shaping matrix matches the source bandwidth.}
    \label{fig:beam_shaping_srandom}
\end{figure}

To corroborate the consistency of the broadband matrix enhancement factor discussed in the main text, Figure~\ref{fig:beam_shaping_srandom} shows the performance when generating a random amplitude and phase superposition of all modes ($\delta\lambda_T \approx \Lambda$,). The results follow the exact same trend as the smiley/sad pattern, confirming the $\sqrt{2}$ bandwidth enhancement factor when using MTM$_{2\Lambda}$.

\subsubsection*{Broadband beam shaping ($\Delta\lambda_S=30$~nm) with MTM($\varphi_{1,2,3}$) }

\begin{figure}[htbp]
    \centering
    \includegraphics[width=0.95\linewidth]{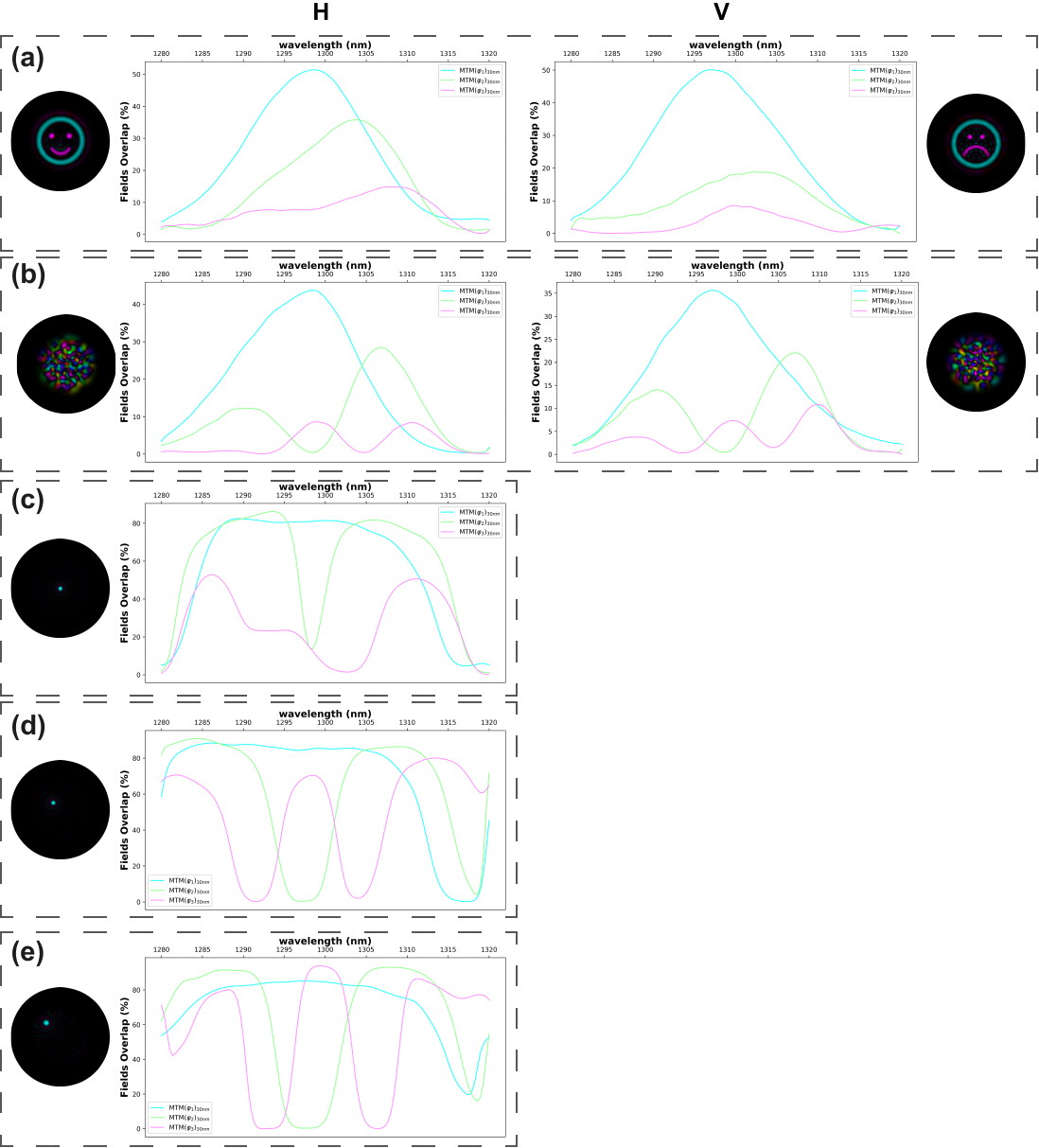}
    \caption{Beam shaping at $\Delta\lambda_S=30$~nm using MTM($\varphi_1$), MTM($\varphi_2$), and MTM($\varphi_3$). Rows: smiley/sad face, random superposition, and off-centre diffraction-limited spots at three radial positions. $\varphi_1$ controls core-weighted patterns; $\varphi_2$ and $\varphi_3$ improve sideband and near-cladding spots that rely on higher-order, shorter-$\Lambda$ modes.}
    \label{fig:beam_shaping_second}
\end{figure}

Figure~\ref{fig:beam_shaping_second} explores the practical utility of the secondary eigenstates by performing beam shaping (smiley/sad, random, and off-centre spots) using MTM($\varphi_{1,2,3}$) at $\Delta\lambda_S=30$~nm. The results reveal a clear spatial segregation of control. MTM($\varphi_1$) heavily favours centre-band control (lower-order modes), while MTM($\varphi_2$) and MTM($\varphi_3$) control the sidebands and higher-order modes. Notably, when attempting to focus a spot near the cladding (which requires higher-order modes with shorter correlation bandwidths), the performance of MTM($\varphi_3$) improves relative to its performance on centre-focused targets, confirming that these secondary eigenstates capture the specific modal information that was "lost" from the primary eigenstate due to spatial decoherence.

% Force all floats before references
\FloatBarrier
\clearpage

\ifdefined\siloadedbymain
\unvbox\sibibbox
\else

\fi

\end{document}